\documentclass[prl,aps,twocolumn,showpacs,preprintnumbers,amsmath,amssymb]{revtex4}
\usepackage{epsfig}
\usepackage{color}
\usepackage{ifthen}
\usepackage{graphicx}
\usepackage{amsmath}
\usepackage{amssymb}

\usepackage{graphicx}
\usepackage{dcolumn}
\usepackage{bm}
\def\Dsl{\hbox{/\kern-.6700em\it D}} 
\def\dsl{\hbox{/\kern-.5300em$\partial$}}

\def\eqa{\begin{eqnarray}}
\def\eeqa{\end{eqnarray}}
\def\eq{\begin{equation}}
\def\eeq{\end{equation}}
\def\be{\begin{equation}}
\def\ee{\end{equation}}
\def\bea{\begin{eqnarray}}
\def\eea{\end{eqnarray}}

\newcommand{\dslash}{\not{\hbox{\kern-2pt $\partial$}}}
\newcommand{\pslash}{\not{\hbox{\kern-2.3pt $p$}}}
 \newtoks\nslashfraction
 \nslashfraction={.13}
 \newcommand{\nslash}[1]{\setbox0\hbox{$ #1 $}
   \setbox0\hbox to \the\nslashfraction\wd0{\hss \box0}/\box0 }

\begin{document}

\preprint{}

\title{Model-Independent Signatures of New Physics in Slow-Roll Inflation}
\author{Mark G. Jackson and Koenraad Schalm}
\affiliation{Instituut-Lorentz for Theoretical Physics, University of Leiden, Leiden 2333CA, The Netherlands}

\date{\today}


\ \newline

\begin{abstract}
\noindent
We compute the universal generic corrections to the power
spectrum in slow-roll inflation due to unknown high-energy physics.
We arrive at this result via a careful integrating out of massive
fields in the ``in-in'' formalism yielding a consistent and predictive
low-energy effective description in time-dependent
backgrounds. The
density power spectrum is universally modified at leading order in $H/M$, the
ratio of the scale of inflation to the scale of new physics;
the tensor power spectrum receives only subleading corrections. In
doing so, we show how to make sense of a physical momentum-cut-off in
loop integrals despite dynamical redshifts, and how the result can be
captured in a combined effective action/effective density matrix,
where the latter contains non-adiabatic terms which modify the boundary conditions.
\end{abstract}

\maketitle

\section{Introduction}
The dawning of the era of precision cosmology demands that we
understand the history of our universe theoretically with the same
accuracy as experiment. The \emph{WMAP} determination of acoustic peaks in
the CMB spectrum to 1\% accuracy has given strong support to the
existence of an era of inflation \cite{wmap7}. Inflation, famously, predicts the
primordial power spectrum of density fluctuations underlying all
structure in the universe. 

This ability of \emph{WMAP} and future {\em Planck} data
\cite{planck} to constrain theoretical models of the early Universe
has set off a scramble to delineate
a theoretically controlled
computation of the primordial inflationary power
spectrum. The textbook approach makes a number of explicit and
implicit assumptions, each of which can affect the power spectrum at
the accuracy measured. The main assumption is that the density
spectrum can be reduced to that of an adiabatic fluid whose
excitations are always weakly coupled to gravity. This can be
conveniently encoded in a single scalar field Lagrangian with
Bunch-Davies initial conditions in a fixed 
inflationary cosmology without metric fluctuations.
The most obvious issue with these assumptions is that the redshifts
implied by the 60 $e$-folds of inflation necessary to solve the horizon
and flatness problem, place the relevant momentum scales in an energy
regime far beyond the Planck-scale, where  in principle gravitational backreaction
and quantum-gravity corrections cannot be ignored
\cite{Brandenberger:1999sw}.  Turning the issue around, it also implies a window of
opportunity that quantum gravity or any other New Physics arising at
high energy scales could have a measurable effect on the power
spectrum. This question was actively pursued some time ago with the
conclusion that in toy models
\cite{Niemeyer:2000eh,Kempf:2000ac,Niemeyer:2001qe,Kempf:2001fa,Martin:2000xs,Brandenberger:2000wr,Brandenberger:2002hs,Martin:2003kp,Easther:2001fi,Easther:2001fz,Easther:2002xe,Kaloper:2002uj,Kaloper:2002cs,Danielsson:2002kx,Danielsson:2002qh,Shankaranarayanan:2002ax,Hassan:2002qk,Goldstein:2002fc,Bozza:2003pr,Alberghi:2003am,Schalm:2004qk,porrati,Porrati:2004dm,Hamann:2008yx,Achucarro:2010da,Starobinsky:2002rp,Chen:2010ef, Ashoorioon:2010xg, Ashoorioon:2004vm}
one can obtain measurable corrections of the order $H/M$, comparable
to intrinsic cosmic variance, with $H$ the Hubble scale and $M$ the
scale of New Physics
\cite{Bergstrom:2002yd,Martin:2003sg,Martin:2004iv,Martin:2004yi,Easther:2004vq,Greene:2005aj,Easther:2005yr,Spergel:2006hy}. 

To truly connect with the data one needs the {\em universal generic}
model-independent corrections to the power spectrum in terms of an
effective field theory that encodes order by order the corrections due
to New Physics at high scales as well as deviations from 
adiabaticity. How to account for New Physics is formally well-understood in terms of Wilsonian effective actions, and for the adiabatic mode this has been
actively pursued recently \cite{Cheung:2007st,Senatore:2010wk,Weinberg:2005vy,Weinberg:2006ac,Weinberg:2008hq,Weinberg:2010wq,vanderMeulen:2007ah,Achucarro:2010da}. However, Wilsonian effective actions are only consistent provided adiabaticity is maintained \cite{Burgess:2002ub,Burgess:2003zw,Burgess:2003hw}. In practice this has meant that energy is assumed to be a conserved quantity. But precisely this is impossible in a cosmological time-dependent background. Redshifts continuously mix the regimes of various scales and strictly speaking a well-defined separation of energy cannot be maintained. This long-standing paradox has fundamentally hampered the construction of low energy effective theories in cosmological spacetimes, {\em literally} since energy is not a conserved quantity.

In a previous letter \cite{Jackson:2010cw} we provided an algorithmic solution to this obstacle to compute the generic new physics corrections to the inflationary power spectrum.
One can generate the universal low energy effective
action by integrating out a massive field in any particular New
Physics model.  This is sensible in a cosmological setting, as long as
one computes late time expectation values directly in non-equilibrium real-time QFT via the Schwinger-Keldysh approach.  One of the
principal difficulties found in previous approaches was the correct procedure to implement
an energy and momentum scale cutoff in a time-dependent background; in intermediate
steps the order of energy and momentum integrals now no longer
commute.  We have overcome this through the realization that the
contributions from the
massive field may be reliably captured in a stationary-phase
approximation for vertex evaluation. This localizes interactions to
fixed moments in time and allows us to unambigously implement the cutoff in momenta.

There are several noteworthy features that this algorithmic solution reveals: \begin{itemize}
\item Most importantly, New Physics and non-adiabatic corrections no longer separate. Integrating out the heavy field not only causes changes in the action, but also in the effective density matrix of the low-energy field. The former can account for non-adiabatic effects. 
\item The non-adiabatic terms are non-local in position space, but localized on a New Physics Hypersurface where the physical momentum ${\bf p}(t)={\bf {k}}/a(t)$ equals the mass of the heavy field $p(t)=M$. Intuitively this is what should happen and was the basis of many earlier \emph{ad hoc} models. Here it is a consequence of integrating out the heavy field. 
\item These terms can be interpreted as modified initial conditions for the low-energy adiabatic inflaton. This again confirms the qualitative insights gained from toy models, but now quantitatively.
\end{itemize}

In this article we apply this cosmological effective field theory approach to the cosmologically relevant scenario of slow-roll inflation.  We compute the corrections to the scalar power spectrum resulting from interactions with a heavy field. Parametrizing the corrections in a manner expedient to comparison with observational data, the {\em universal generic} correction to the scalar power spectrum due to unknown high-energy physics equals
\[ \frac{\Delta P_\zeta}{P}  \sim \frac{H}{M} \left( C_0 + C_1 \epsilon_1 + \cdots  \right) . \]
where $\epsilon_1$ is the first slow-roll coefficient.  The power of the approach, however, is that each of the variables $C_i$ can be {\em computed} in terms of the parameters of whichever theoretical model for the unknown New Physics one has in mind.  For the power spectrum the result is merely an amplitude correction, but it will be very interesting to see the effect in higher-order correlation functions.

This article is organized as follows.  In \S2 we present a simple model of slow-roll inflation containing new physics at high energies and analyze the field fluctuations in two gauges which will prove useful.  In \S3 we review the in-in formalism and how to compute field fluctuation correlations for our theory.  In \S4 we calculate the scalar and tensor power spectra in the spatially flat gauge, then convert these to the uniform density gauge and parametrize the answer in a way which facilitates easy comparison against observation.  In \S5 we discuss observational prospects and conclude.

\section{Single Field Slow-Roll inflation plus a Massive field}

The minimal action description for inflation is a scalar field coupled to gravity,
\begin{equation}
\label{sinf}
 S_{\inf} [\phi] = \int d^4 x \sqrt{g} \left[ \frac{1}{2} M_{\rm pl}^2 R - \frac{1}{2} (\partial \phi)^2 - V_{\rm inf}(\phi) \right] .
\end{equation}
We assume a spatially homogeneous ansatz for the background metric 
\[ ds^2 = -dt^2 + a(t)^2 d {\bf x}^2.  \]
Defining $H \equiv {\dot a} / a$, the background equations of motion are
\begin{eqnarray}
\label{eom1}
{\dot \phi} &=& - \frac{M^2_{\rm pl}}{4 \pi} H' (\phi), \\
\label{eom2}
\left[ H'(\phi) \right]^2 - \frac{12 \pi}{M^2_{\rm pl}} H (\phi)^2 &=& - \frac{32 \pi^2}{M^4_{\rm pl}} V(\phi).
\end{eqnarray}
Primes denote derivatives with respect to the field, while overdots denote derivatives with respect to coordinate time.  

To facilitate analysis of inflating solutions, one defines the (first) Hubble slow-roll parameter $\epsilon_1$
\begin{equation}
\label{sl1}
 \epsilon_1(\phi) \equiv \frac{M^2_{\rm pl}}{4 \pi} \left[ \frac{H'(\phi)}{H(\phi)} \right]^2. 
 \end{equation}
Inflation occurs for $\epsilon_1 < 1$ and ends when $\epsilon_1 = 1$, which we use to define $t_{\rm
  end}$.  
It will prove useful to also define a second slow-roll parameter,
\[ \epsilon_{2}(\phi) \equiv \frac{ d \ln  \epsilon_1}{d N} = \frac{M^2_{\rm pl} H''(\phi)}{4 \pi H(\phi)} . \]
One could continue defining an infinite hierarchy of such parameters $\epsilon_n$, but this will be sufficient for our purposes.

Of course there is a direct relationship between $V_{\rm inf}(\phi)$ and $\epsilon_1, \epsilon_2$.  This is given by
\begin{eqnarray*}
 V(\phi) &=& \frac{3 M^2_{\rm pl}}{8 \pi^2} H^2 \left( 1 - \frac{\epsilon_1}{3} \right), \\
 \frac{d V}{d \phi} &=& - 3H^2 M_{\rm pl} \sqrt{ \frac{\epsilon_1}{4 \pi}} \left(1 - \frac{\epsilon_1}{3} + \frac{\epsilon_2}{6} \right), \\
 \frac{d^2 V}{d \phi^2} &=& 3H^2 \left( 2 \epsilon_1 - \frac{\epsilon_2}{2} \right).
\end{eqnarray*}
It will be helpful to define the so-called ``potential slow- roll" parameters
\[ \epsilon_V(\phi) \equiv \frac{M^2_{\rm pl}}{16 \pi} \left( \frac{V'}{V} \right)^2, \hspace{0.5in} \eta_V(\phi) \equiv \frac{M^2_{\rm pl}}{8 \pi} \frac{V''}{V} . \]
We can then invert this to express $\epsilon_1, \epsilon_2$ in terms of $\epsilon_V, \eta_V$, with exact solutions possible in the case of power-law inflation.
\subsection{Perturbative Interactions}
We may take the initial value of the inflaton to be $\phi(t_{\rm in}) = 0$, so that the potential near this point is approximated as
\[ V_{\inf} (\phi) \approx V_0 + j \phi + \frac{m^2}{2} \phi^2 + \frac{g_0}{3!} \phi^3 + \cdots  \]
where $V_0, j, m^2, g_0$ are all constants.  Thus $\epsilon_V, \eta_V$, and hence $\epsilon_1, \epsilon_2$, can be solved for in terms of these coupling constants.

To the inflationary action (\ref{sinf}) we then add a massive field $\chi$ with renormalizable interactions to the inflaton:
\begin{eqnarray*}
\label{action}
S_{\rm new}[\phi,\chi] &=& - \int d^4 x \sqrt{g} \left[  \frac{1}{2} (\partial \chi)^2  + \frac{1}{2} M^2 \chi^2  \right. \\
&& \left. \hspace{1in} + \frac{g_1}{2} \phi^2 \chi + \frac{g_2}{2} \phi \chi^2 \right].
 \end{eqnarray*}
We will be working to cubic order in field fluctuations, but are only interested in single contractions of the heavy field fluctuations, justifying our expansion in $\chi$ to second order.   The combined action $S \equiv S_{\rm inf} + S_{\rm new}$ will then produce an inflationary period but contain New Physics at scales determined by the couplings $g_i$ and mass scale $M$.  
 
 A naive approach, based on static background quantum field theory intuition, might be the following: since we are presumed to be below the energy scale where $\chi$-quanta can be created, we can simply integrate out this field to yield the following effective `New Physics' potential for $\phi$:
\begin{eqnarray*}
V_{\rm new}(\phi) &=& - \frac{g_1^2}{4} \phi^2 \frac{1}{\Box - M^2} \phi^2 \\
&\approx& \frac{g^2_1}{4M^2} \phi^4 + \cdots \hspace{0.5in} {\rm (naive)}. 
\end{eqnarray*}
 This is incorrect, for a variety of reasons.  In the remainder of this article we will delineate the correct procedure to obtain the effective description.
 
\subsection{Fluctuations}
Let us now assume that the spatially-homogeneous component ${\phi}_{0}(t)$ satisfies its own equations of motion to lowest order in $g_i$ and consider fluctuations around this background \footnote{The interactions are tadpole terms, and we are therefore not expanding about a saddle point of the total action.  However for the purposes of deriving a low energy Wilsonian action this is not an obstacle, as we address later in detail.},
\[ \varphi(t, {\bf x}) \equiv {\phi}(t, {\bf x}) - {\phi}_{0}(t) . \]
Although our observable of interest is the $\varphi$-power spectrum
which does not usually require gauge-fixing, we wish to incorporate
the effects of interactions for which the inflaton field and metric
fluctuations mix together, requiring the construction of
gauge-invariant quantities.  We therefore also consider tensor fluctuations.  Utilizing the ADM formalism, we parametrize the metric as
\[ ds^2 = -N^2 dt^2 + h_{ij} (dx^i + N^i dt)(dx^j + N^j dt)  \]
with $N$ the lapse function and $N^i$ the shift vector. Substituting this into the action produces
\begin{eqnarray}
\nonumber
S &=& \int dt d^3 {\bf x} \sqrt{h} N \left[ \frac{M^2_{\rm pl}}{2} {\bf R}  + \frac{M^2_{\rm pl}}{2N^2} (E_{ij} E^{ij} - E^2 ) \right. \\
\label{admaction}
&& \hspace{0.2in} + \frac{( {\dot \phi} - N^i \partial_i \phi)^2}{2N^2} - \frac{1}{2} h^{ij} \partial_i \phi \partial_j \phi \\ 
\nonumber
&& \left.  \hspace{0.2in} + \frac{( {\dot \chi} - N^i \partial_i \chi)^2}{2N^2} - \frac{1}{2} h^{ij} \partial_i \chi \partial_j \chi  - V(\phi,\chi)  \right] 
\end{eqnarray}
where 
\begin{eqnarray*}
h &\equiv& \det h_{ij}, \\
{\bf R} &=& {\rm Ricci \ curvature \ of \ spatial \ metric}, \\
E_{ij} &\equiv& \frac{1}{2} \left( {\dot h}_{ij} - \nabla_i N_j - \nabla_j N_i \right), \\
E &=& {E_i}^i.
\end{eqnarray*}
The equations of motion for $N$ and $N^i$ are just the Hamiltonian and momenta constraints,
\begin{eqnarray}
\label{eomn}
0 &=& \frac{M^2_{\rm pl}}{2} {\bf R} - \frac{M^2_{\rm pl}}{2N^2} (E_{ij} E^{ij} - E^2 ) - \frac{({\dot \phi} - N^i \partial_i \phi)^2}{2N^2}  \\
\nonumber
&&  - \frac{1}{2} h^{ij} \partial_i \phi \partial_j \phi  -  \frac{({\dot \chi} - N^i \partial_i \chi)^2}{2N^2}- \frac{1}{2} h^{ij} \partial_i \chi \partial_j \chi  - V, \\
\label{eomni}
0 &=& M^2_{\rm pl} \nabla_j \left[ \frac{ {E_j}^i - {\delta_j}^i E}{N} \right] \\
\nonumber
&& - \frac{\partial_i \phi ( {\dot \phi} - N^j \partial_j \phi)}{N} -  \frac{\partial_i \chi ( {\dot \chi} - N^j \partial_j \chi)}{N}. 
\end{eqnarray}
In solving for $N$ and $N^i$ we can ignore the effect of $\chi$; as we assume a single-field inflation model, rather the the hybrid-inflation model of $\phi$ and $\chi$.  This is valid because the corrections to the free equations of motion will be $\mathcal O(g^2)$.  Since this is the same order as the fluctuations we are computing, we may trust our unperturbed background solution.  

There are two choices of gauge we will employ: spatially flat and uniform density.  That is, using different gauges we may exchange perturbations in $\phi$ for (certain) perturbations in $g_{\mu \nu}$ and vice-versa.  Both will be useful at different points in our calculation.
\subsubsection{Spatially Flat Gauge}
We will perform the interaction calculations inside the horizon using the spatially flat gauge.  This corresponds to the choice 
\begin{eqnarray*}
h_{ij} &=& a(t)^2 {\hat h}_{ij}, \hspace{0.3in}  {\hat h}_{ij} = \delta_{ij} + {\gamma}_{ij} , \\
 \partial_i {\gamma}_{ij} &=& 0, \hspace{0.7in} {\gamma}_{ii} = 0.
\end{eqnarray*}
Fluctuations are then parametrized by $\varphi, {\gamma}_{ij}$ and $\chi$.  We now need the solution for $N,N^i$ in terms of these fluctuations.  Since we are interested in quadratic fluctuations of $\varphi$ and ${\gamma}_{ij}$, it suffices to compute the background corrections to first order.  This is because any third order terms for the background would multiply the zero-order terms for the fields, which are automatically satisfied.  We denote $N = 1 + N_1$ so that $N_1$ is a first-order perturbation, as is $N^i$. The solutions are given by
\begin{eqnarray*}
N &=& 1+ \sqrt{ \frac{\epsilon_1}{2}} \frac{\varphi}{M_{\rm pl}}, \\
N_i &=& \partial_i n, \hspace{0.5in} \partial_i \partial^i n = - \frac{ \epsilon_1}{ \sqrt{2}} \frac{d}{dt} \left( \frac{1}{ \sqrt{\epsilon_1}} \frac{\varphi}{M_{\rm pl}} \right).
\end{eqnarray*}
Substituting this back into the action (\ref{admaction}) yields (to lowest relevant order)
\begin{eqnarray*}
S &=& - \int dt d^3 {\bf x} \ a^3 \left[ - \frac{M^2_{\rm pl}}{2} \left( \partial {\gamma}_{ij} \right)^2 + \frac{1}{2} (\partial \varphi)^2 + j \varphi \right. \\
&+& \left( m^2 - \frac{2}{a^3} \frac{d}{dt} \left( a^3 H \epsilon_1 \right) \right) \varphi^2 + \frac{1}{2} (\partial \chi)^2 + \frac{1}{2} M^2 \chi^2 \\
&+& \left. \frac{g_1}{2} \varphi^2 \chi + \frac{g_2}{2} \varphi \chi^2 + \frac{1}{2} \gamma_{ij} \partial_i \chi \partial_j \chi \right]. 
\end{eqnarray*}
Note the appearance of a small inflaton mass induced from the gauge-fixing.  In writing this we have omitted the $ \varphi \chi$ mass-mixing term which is subleading in $H/M$. Additionally, we have neglected all interactions which do not contain $\chi$, since we are only interested in the effect of this heavy field.   One of these neglected terms, the $h^{ij} \partial_i \varphi \partial_j \varphi$ coupling, produces an $M_{\rm pl}$-scale interaction and it would be interesting to compare this to the result obtained here.  We save this for a future study.
\subsubsection{Uniform Density Gauge}
At the moment of scalar fluctuation horizon crossing we then convert these to the uniform density gauge since these will then stay constant and are good observables.  This gauge corresponds to the choice 
\begin{eqnarray*}
\varphi &=& 0, \hspace{0.4in} h_{ij} = a(t)^2 \left[ (1+2 \zeta)\delta_{ij} + \gamma_{ij} \right], \\
 \partial_i {\gamma}_{ij} &=& 0, \hspace{0.4in} {\gamma}_{ii} = 0.
\end{eqnarray*}
Note that the $\chi$ and $\gamma$ fluctuations remain identical, while scalar fluctuations are now parametrized by $\zeta$.  The conversion between gauges is 
\begin{equation}
\label{gaugeconversion}
 \zeta = \sqrt{ \frac{4 \pi}{\epsilon_1}} \frac{\varphi}{M_{\rm pl}}. 
 \end{equation}
Of course, were we to include the effects of $\chi$ on the background evolution, this hybrid-inflation model would \emph{not} have constant scalar fluctuations.  But as addressed previously, the leading-order $\chi$-fluctuations employ a single-field background and hence these fluctuations will still have constant superhorizon $\zeta$. 
\section{The In-In Formalism}
\subsection{In-Out Amplitudes Versus In-In Expectation Values}
Quantum field theory in a static background most often employs the ``in-out'' formalism to produce scattering amplitudes.  Defining the states $| {\rm in} \rangle$ and $| {\rm out} \rangle$ in the asymptotic past and future, respectively, amplitudes are defined as
\[ \mathcal A_{\mathcal O} \sim \langle {\rm out} | \mathcal O | {\rm
  in} \rangle . \]
Cross-sections are then obtained by squaring the amplitude.
For non-equilibrium systems, such as a cosmological background, the
fundamentally sound approach to computing expectation values such as
the power spectrum is the Schwinger-Keldysh approach
\cite{Calzetta:1986cq}.  

The procedure is the following.  At some early time $t_{\rm in}$ (in the present context, the onset of inflation) we begin with a pure state $| {\rm in}(t_{\rm in}) \rangle$, then evolve the system for the bra- and ket-state separately until some late time $t$, when we evaluate the expectation value:
\begin{eqnarray}
\label{inincorrelation}
\langle \mathcal O(t) \rangle &\equiv& \langle {\rm in}(t) | \mathcal O(t) | {\rm in} (t) \rangle \\
\nonumber
&=&  \langle {\rm in}(t_{\rm in} ) | e^{i \int _{t_{\rm in}} ^t  dt' H(t')} \mathcal O(t) e^{-i \int _{t_{\rm in}} ^t  dt'' H(t'')} | {\rm in} (t_{\rm in}) \rangle .
\end{eqnarray}
Traditionally, the in-state $| {\rm in} \rangle$ is taken to be the Bunch-Davies vacuum state \cite{Bunch:1978yq}, but this is not necessarily so.  Expanding cosmological backgrounds allow for a more general class of vacua, which can be heuristically considered to be excited states of inflaton fluctuations.  In the present context, we will find that integrating out high-energy physics generically results in boundary terms in the effective action, which represent such excited states.

If we denote the fields representing the ``evolving'' ket to be $\{ \varphi_+, \chi_+, {\gamma}_{ij,+} \}$ and those for the ``devolving'' bra to be $\{ \varphi_-, \chi_- , {\gamma}_{ij,-} \}$, the in-in expectation value (\ref{inincorrelation}) can be computed from the action
\begin{equation}
\label{keldyshaction}
 \mathcal S \equiv S[\varphi_+, \chi_+,{\gamma}_{ij,+}] - S[\varphi_-, \chi_-,{\gamma}_{ij,-}] . 
 \end{equation}
together with the constraint that $\varphi_+(t)=\varphi_-(t)$, $\chi_+(t)=\chi_-(t)$ and ${\gamma}_{ij,+}(t)={\gamma}_{ij,-}(t)$.
It is then helpful to transform into the Keldysh basis,
\begin{eqnarray*}\
\nonumber
 {\bar \varphi} &\equiv& (\varphi_+ + \varphi_-)/2,  \hspace{0.5in} {\Phi} \equiv \varphi_+ - \varphi_-, \\
{\bar \chi} &\equiv& (\chi_+ + \chi_-)/2,  \hspace{0.5in} {\rm X} \equiv \chi_+ - \chi_-, \\
{\bar {\gamma}}_{ij} &\equiv& ({\gamma}_{ij,+} + {\gamma}_{ij,-})/2,  \hspace{0.3in} {\rm \Gamma} \equiv {\gamma}_{ij,+} - {\gamma}_{ij,-}.
\end{eqnarray*}
In this basis the total action (\ref{keldyshaction}) equals
\begin{eqnarray}
\nonumber
&& \mathcal S [ {\bar \varphi}, {\Phi}, {\bar \chi},  {\rm X}, {\bar {\gamma}}, \Gamma] = \\
\label{skeld}
\nonumber
&& - \int d^3 {\bf x} dt \ a(t)^3 \left[ - M^2_{\rm pl} \partial {\bar \gamma}_{ij} \partial {\Gamma}_{ij} + \partial {\bar \varphi} \partial \Phi + j \Phi  \right. \\
\nonumber
&& + \left( m^2 - \frac{2}{a^3} \frac{d}{dt} \left( a^3 H \epsilon_1 \right) \right) {\bar \varphi} \Phi +  \partial {\bar \chi} \partial {\rm X} + M^2  {\bar \chi} {\rm X} \\
\nonumber
&&   + g_1 {\bar \varphi} \Phi {\bar \chi} + \frac{g_1}{2} \left( {\bar \varphi}^2  + \frac{  {\Phi}^2}{4} \right) {\rm X} +  g_2 {\bar \chi} {\rm X} {\bar \varphi}+  \frac{g_2}{2} \left( {\bar \chi}^2  + \frac{  {\rm X}^2}{4} \right) {\Phi}  \\
\label{keldyshinteractions}
&& \left. + {\bar {\gamma}}_{ij} \partial_i {\bar \chi} \partial_j {\rm X} + \frac{1}{2} \Gamma_{ij} \left( \partial_i {\bar \chi} \partial_j {\bar \chi} + \frac{1}{4} \partial_i {\rm X} \partial_j {\rm X} \right) \right].
\end{eqnarray}
\subsection{Density Matrices}
Although we started with a pure state in (\ref{inincorrelation}), in general we could take expectation values with respect to a mixed state,
\[ \langle \mathcal O (t) \rangle = \sum_{i,j} \langle {\rm in}_i (t)| \mathcal O(t) | {\rm in}_j (t) \rangle \rho_{ij} . \]
Here the density matrix $\rho$ is normalized so that ${\rm Tr} \rho = \sum_i \rho_{i,i} = 1$.  In the path integral language, the density matrix is equal to the logarithm of the imaginary component of the action, so that
\[ \mathcal S = {\rm Re} \mathcal S - i \ln \rho . \]
We will find that integrating out $\chi$ will generically lead to a mixed state for $\varphi$ and $\gamma$, since we are removing states from a unitary process.
\subsection{Perturbative Solution of Fluctuations}
\begin{figure}
\begin{center}
\includegraphics[width=3.5in]{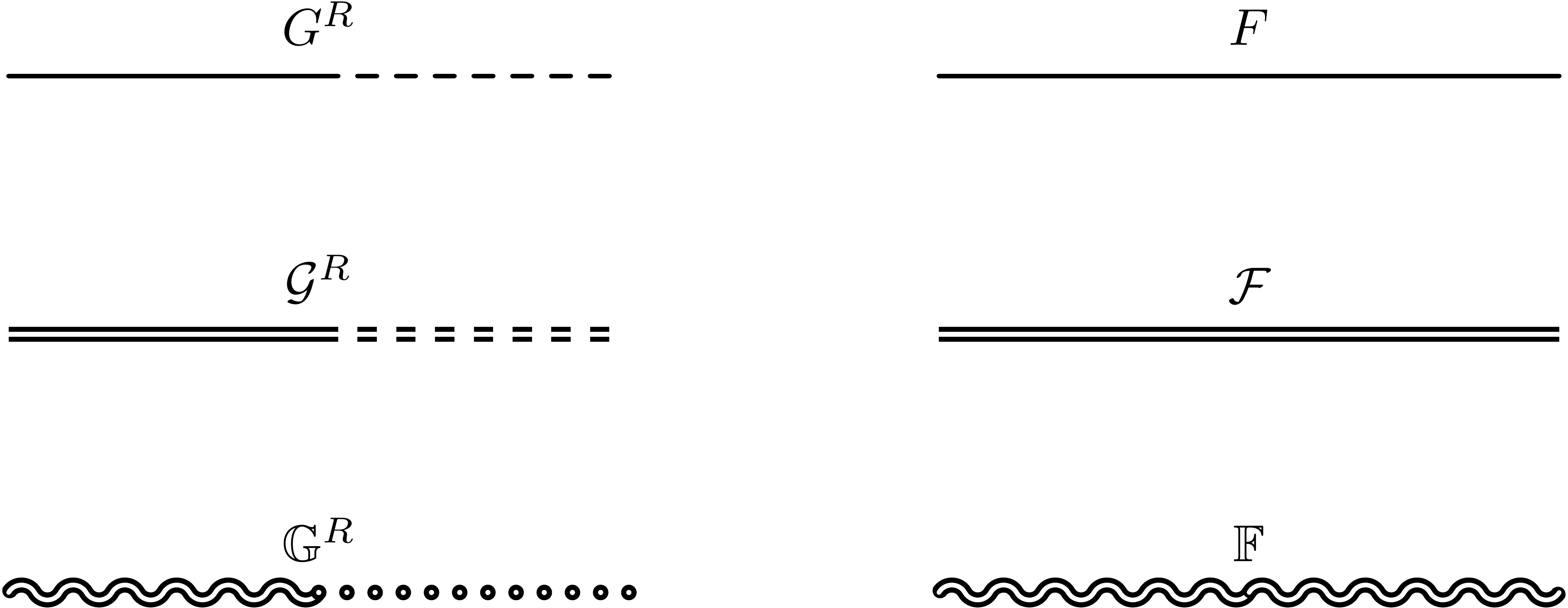} \\
\vspace{0.2in}
\parbox{28mm}{\includegraphics[scale=0.25]{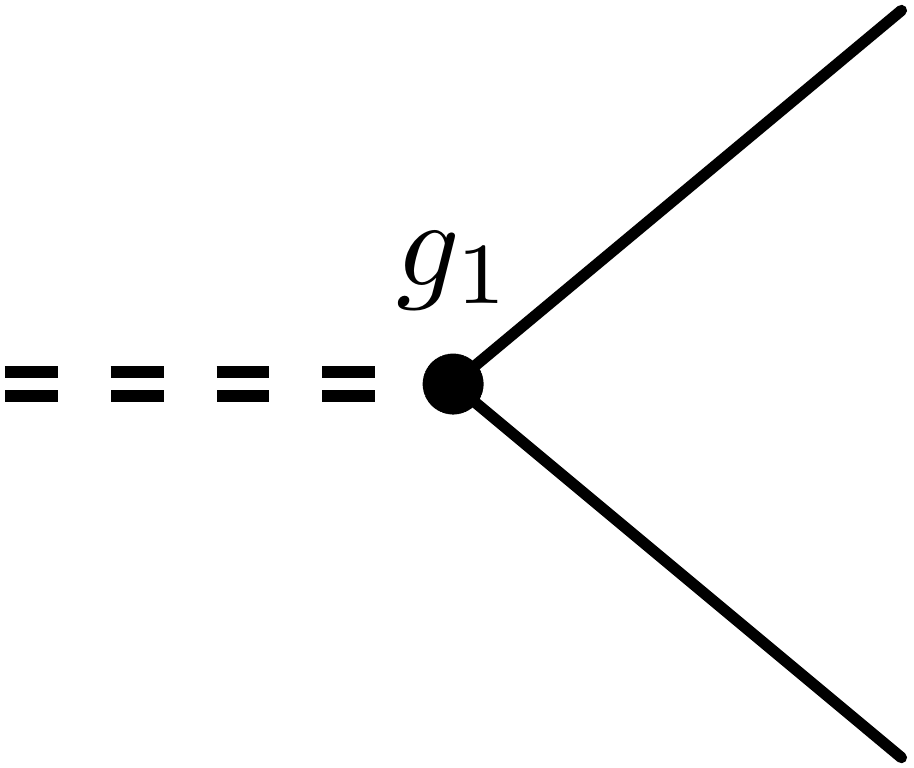}}
\parbox{28mm}{\includegraphics[scale=0.25]{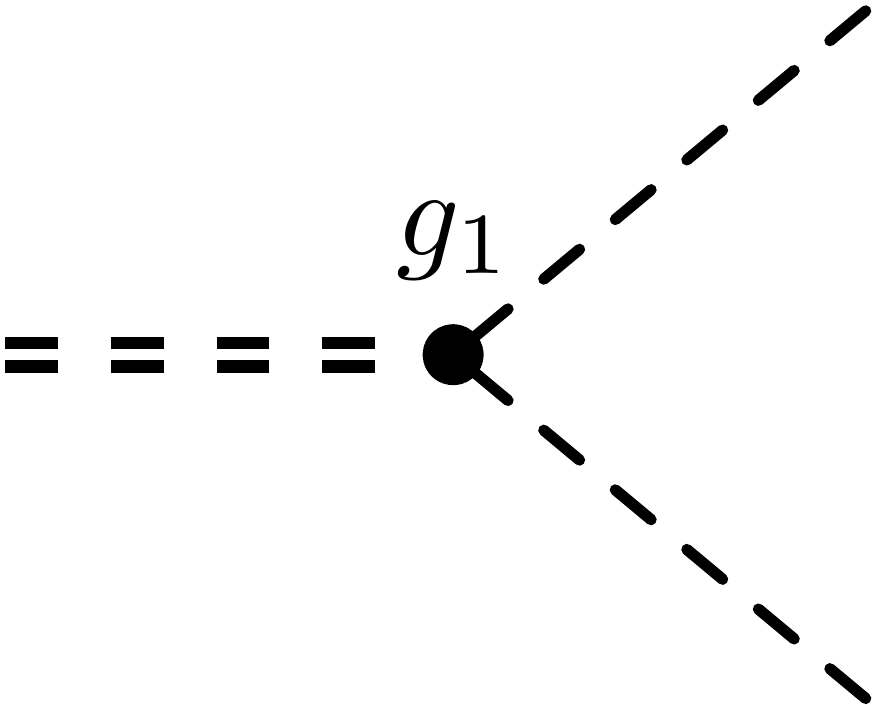}}
\parbox{28mm}{\includegraphics[scale=0.25]{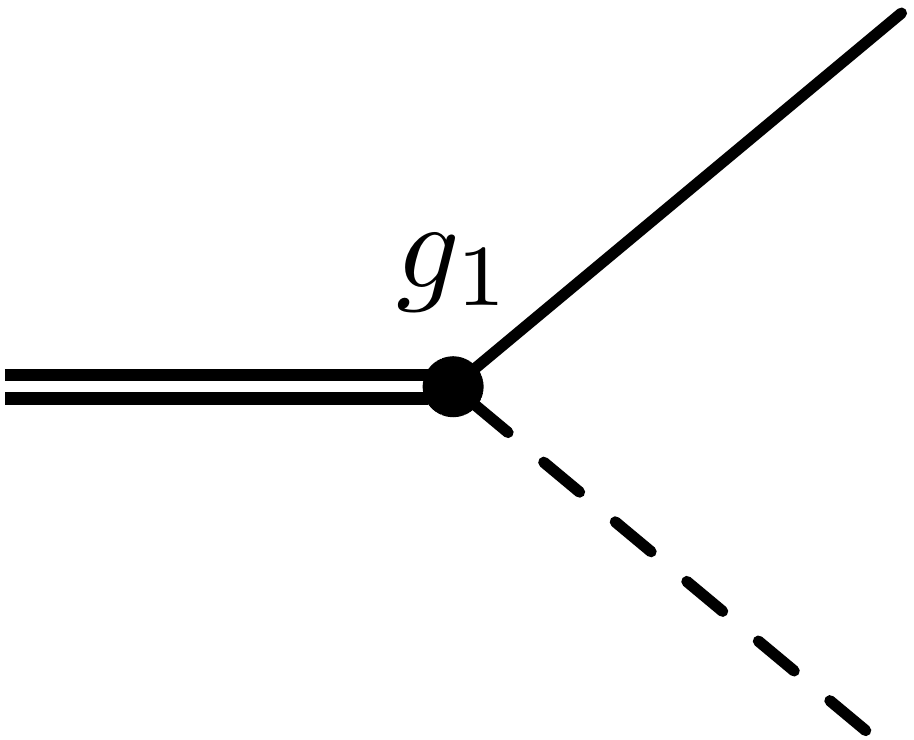}} \\
\vspace{0.1in}
\parbox{28mm}{\includegraphics[scale=0.25]{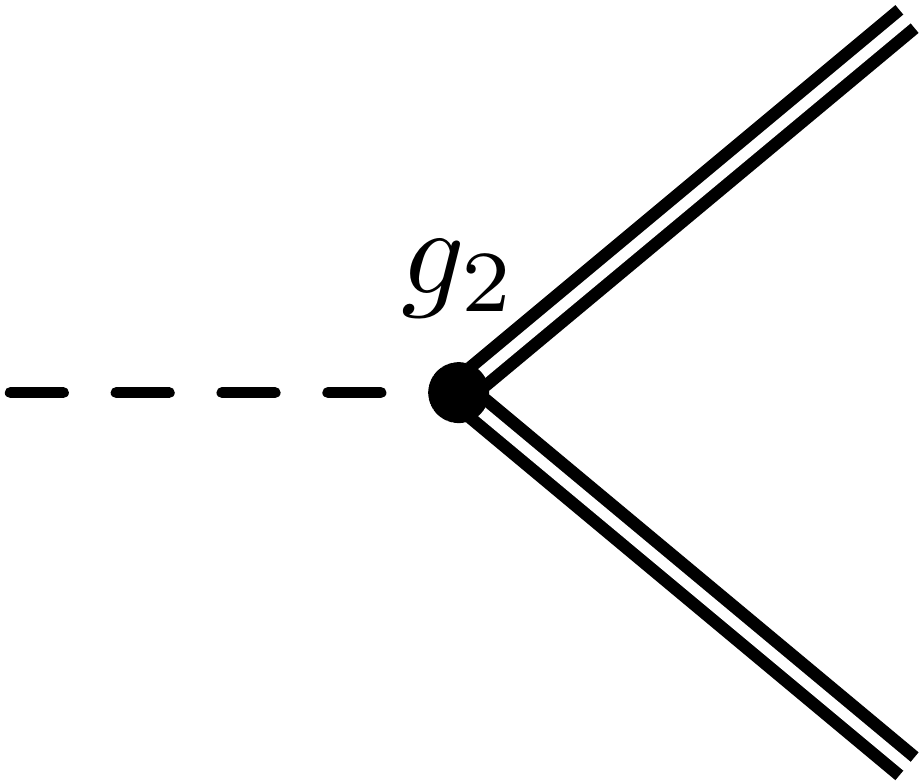}}
\parbox{28mm}{\includegraphics[scale=0.25]{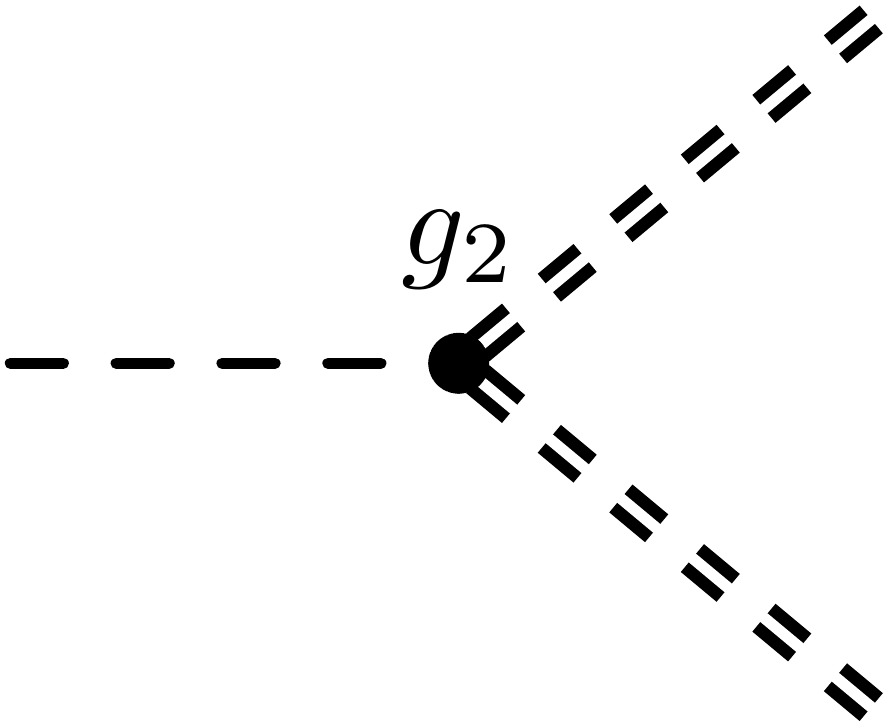}}
\parbox{28mm}{\includegraphics[scale=0.25]{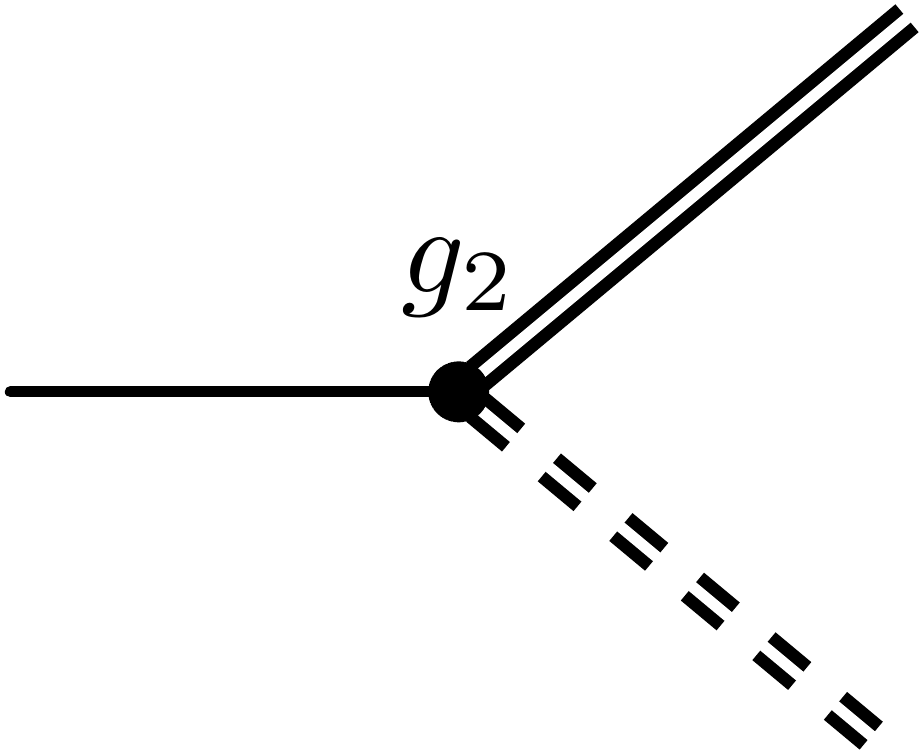}} \\
\vspace{0.1in}
\parbox{28mm}{\includegraphics[scale=0.25]{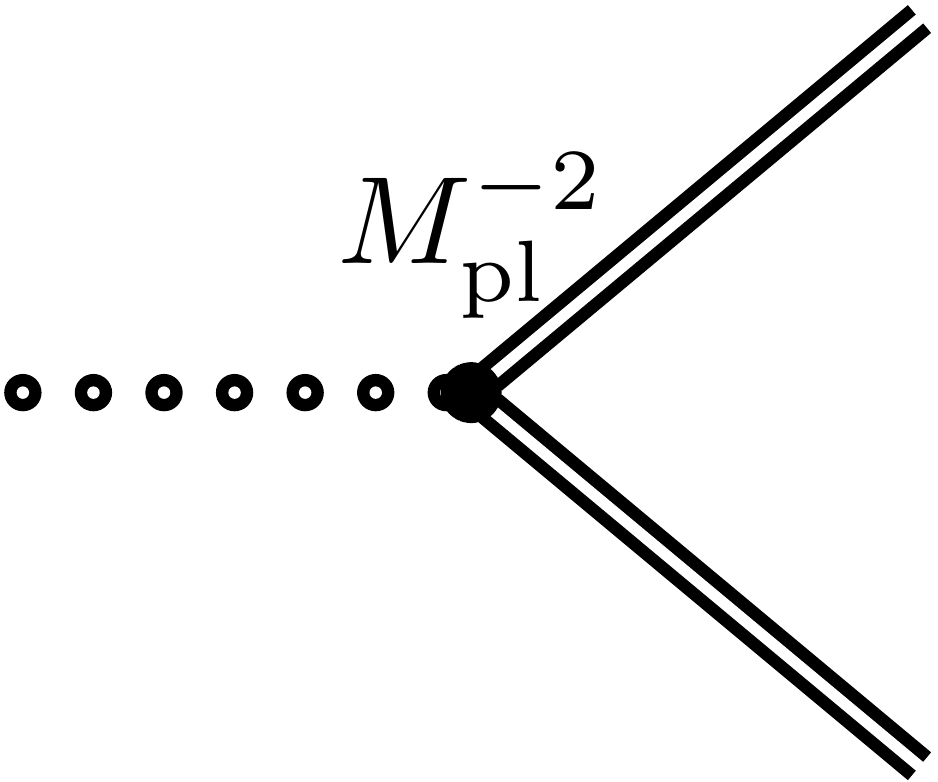}}
\parbox{28mm}{\includegraphics[scale=0.25]{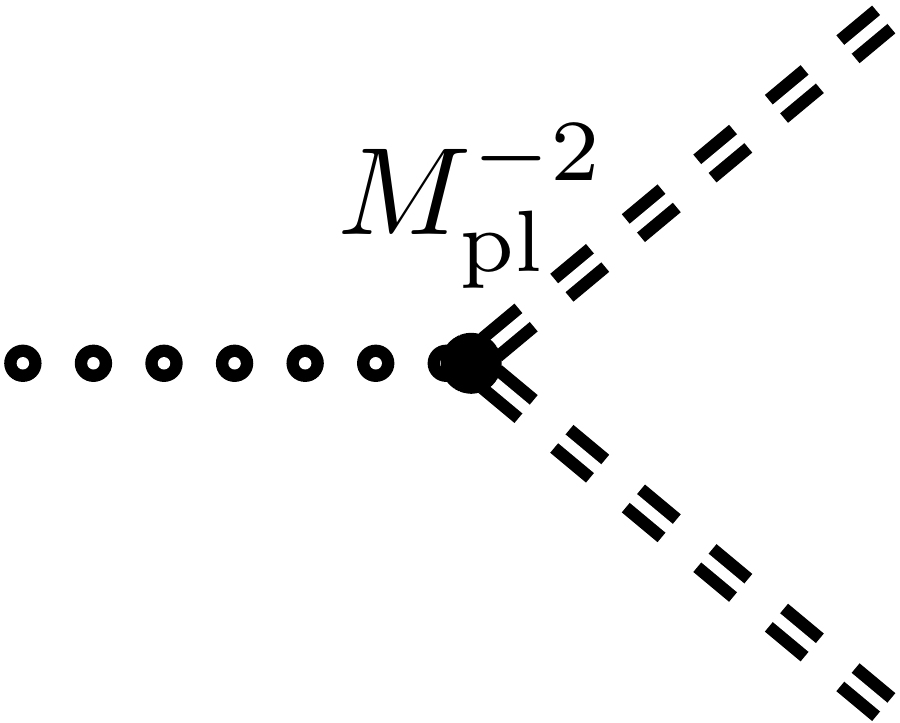}}
\parbox{28mm}{\includegraphics[scale=0.25]{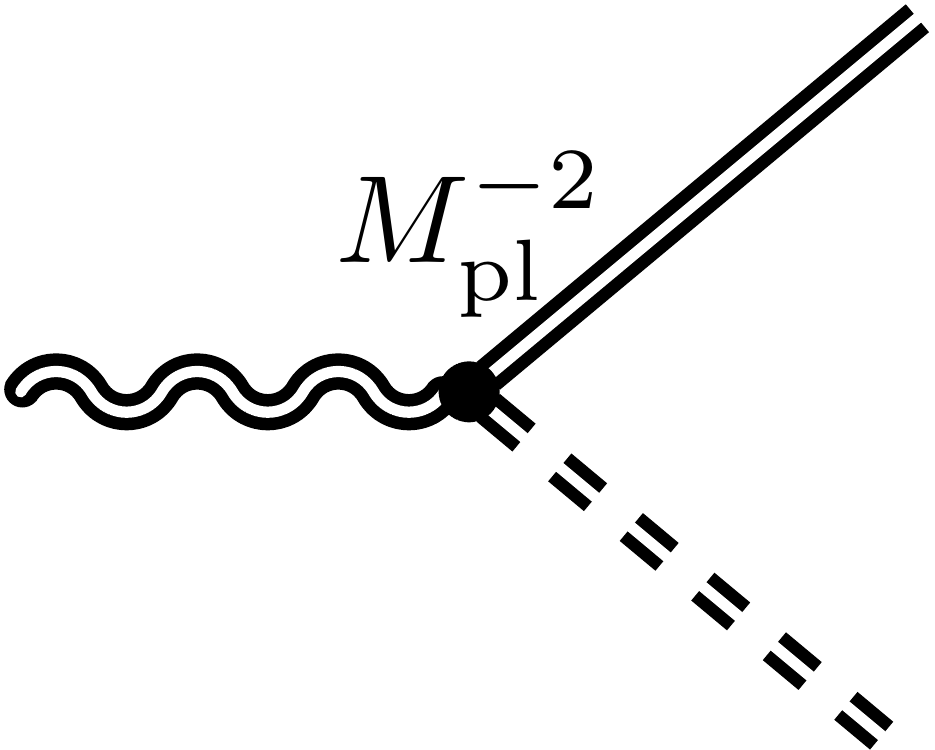}}
\caption{Feynman rules for our theory in the in-in formalism.  Single solid lines indicate contractions of ${\bar \varphi}$, dashed single lines indicate those of $\Phi$, with analogous notation for double lines indicating the heavy field components $\{ {\bar \chi},{\rm X} \}$, wiggly lines indicating the graviton component ${\bar {\gamma}}$ and circled dots for $ \Gamma $.  }
\label{feynman}
\end{center}
\end{figure}

The equations of motion for the fluctuations are
\begin{eqnarray}
\nonumber
&& \hspace{-0.5in} \partial^2 {\bar \varphi} + 3 H {\dot {\bar \varphi}} - j - \left( m^2  - \frac{2}{a^3} \frac{d}{dt} \left( a^3 H \epsilon_1 \right) \right) {\bar \varphi} = g_1 {\bar \varphi} {\bar \chi} + \frac{g_1}{4} \Phi {\rm X}, \\
\nonumber
&& \partial^2 {\bar \chi} + 3H {\dot {\bar \chi}} - M^2 {\bar \chi} = \frac{g_1}{2} \left( {\bar \varphi}^2 + \frac{\Phi^2}{4} \right) + \frac{g_2}{4} \Phi {\rm X} \\
\nonumber
&& \hspace{1.3in} - \partial_j \left( {\bar \gamma}_{ij} \partial_i \chi + \frac{1}{4} \Gamma_{ij} \partial_i \chi \right), \\
\label{perts}
&&  \partial^2 {\bar {\gamma}}_{ij} +3H {\dot {\bar {\gamma}}}_{ij} = \frac{M^{-2}_{\rm pl}}{2} \left( \partial_i {\bar \chi} \partial_j {\bar \chi} + \frac{1}{4} \partial_i {\rm X} \partial_j {\rm X} \right).
\end{eqnarray} 
We will now solve these perturbatively.  The Feynman rules are summarized by Figure~\ref{feynman}, where we have used the same diagrammatic notation as in \cite{vanderMeulen:2007ah}.

\subsubsection{Zeroth Order}
We first consider the free fluctuation equations of motion.  To do this it is helpful to switch to conformal time $\tau$ defined by
\[ ds^2 = a(\tau)^2 (- d\tau^2 + d {\bf x}^2). \]
Neglecting the interactions in equations (\ref{perts}), then Fourier transforming into the comoving momentum basis, they become
\begin{eqnarray*}
{\bar \varphi}_{\bf k}^{(0) \prime \prime} + 2 \mathcal H {\bar \varphi}_{\bf k}^{(0) \prime} - \left( k^2 + m^2 a^2 - \frac{2}{a^3} \frac{d}{dt} \left( a^3 H \epsilon_1 \right) \right) {\bar \varphi}_{\bf k}^{(0) } &=& 0, \\
{\bar \chi}_{\bf k}^{(0) \prime \prime} + 2 \mathcal H {\bar \chi}_{\bf k}^{(0) \prime} - \left( k^2 + M^2 a^2 \right) {\bar \chi}_{\bf k}^{(0)}  &=& 0, \\
 {\bar {\gamma}}^{(0)\prime \prime}_{ij,\bf k} + 2 \mathcal H  {\bar {\gamma}}^{(0) \prime}_{ij, \bf k} - k^2 {\bar {\gamma}}^{(0) }_{ij,\bf k} &=& 0.
\end{eqnarray*}
Here we have introduced the conformal Hubble parameter $\mathcal H (\tau) \equiv a'/a^2$ and suppressed all indices on ${\bar \gamma}$.  The solution for ${\bar \varphi}_{\bf k}^{(0)}$ is given by the Hankel function of the first kind,
\begin{eqnarray}
\nonumber
U_{\bf k}(\tau) &=&  - \frac{ \sqrt{- \pi \tau}}{2a(\tau)} H_\nu^{(1)}(-k \tau), \\
\label{varphisol}
  \nu &\approx& \frac{3}{2} + \epsilon_1 + \frac{\epsilon_2}{2}.
 \end{eqnarray}
If we expand in terms of the slow-roll parameters, this is the familiar expression
\[ U_{\bf k}(\tau) \approx \frac{H}{ \sqrt{2k^3}} (1 - ik \tau) e^{-ik \tau} + \cdots \]
The linearly independent solution $U^*_{\bf k}(\tau)$ is simply the complex conjugate and hence uses $H_\nu^{(2)}$. 

For the massive $\chi$-fluctuations the free field solution is also a Hankel function but can written more transparently using the WKB approximation as  
\begin{equation}
\label{chisol}
V_{\bf k} (\tau) \approx \frac{1}{a(\tau)} \frac{\exp \left[ - i \int ^\tau_{\tau_{\rm in}} d \tau' \sqrt{  k^2 + \frac{M^2}{H(\tau')^2 \tau^{\prime 2}}} \right]}{\sqrt{ 2} \left( k^2 + \frac{M^2}{H(\tau)^2 \tau^2} \right)^{1/4} } .
\end{equation}
The time-dependent frequency here is $\omega(\tau)~\equiv~\sqrt{  k^2 + \frac{M^2}{H(\tau)^2 \tau^2}}$, and the WKB approximation $| \dot{\omega}|/\omega^2~\ll~1$ is always valid for $H/M \ll 1$. 

For ${\bar \gamma}_{ij}$-fluctuations the free solution is again given by Hankel function but of a different order,
\begin{eqnarray*}
W_{\bf k}(\tau) &=&  - \frac{ \sqrt{- \pi \tau}}{2a(\tau)} H_\mu^{(1)}(-k \tau), \hspace{0.4in} \mu \approx \frac{3}{2} +\epsilon_1.
\end{eqnarray*}

\subsubsection{Higher Order}
\begin{figure}
\begin{center}
\includegraphics[scale=0.2]{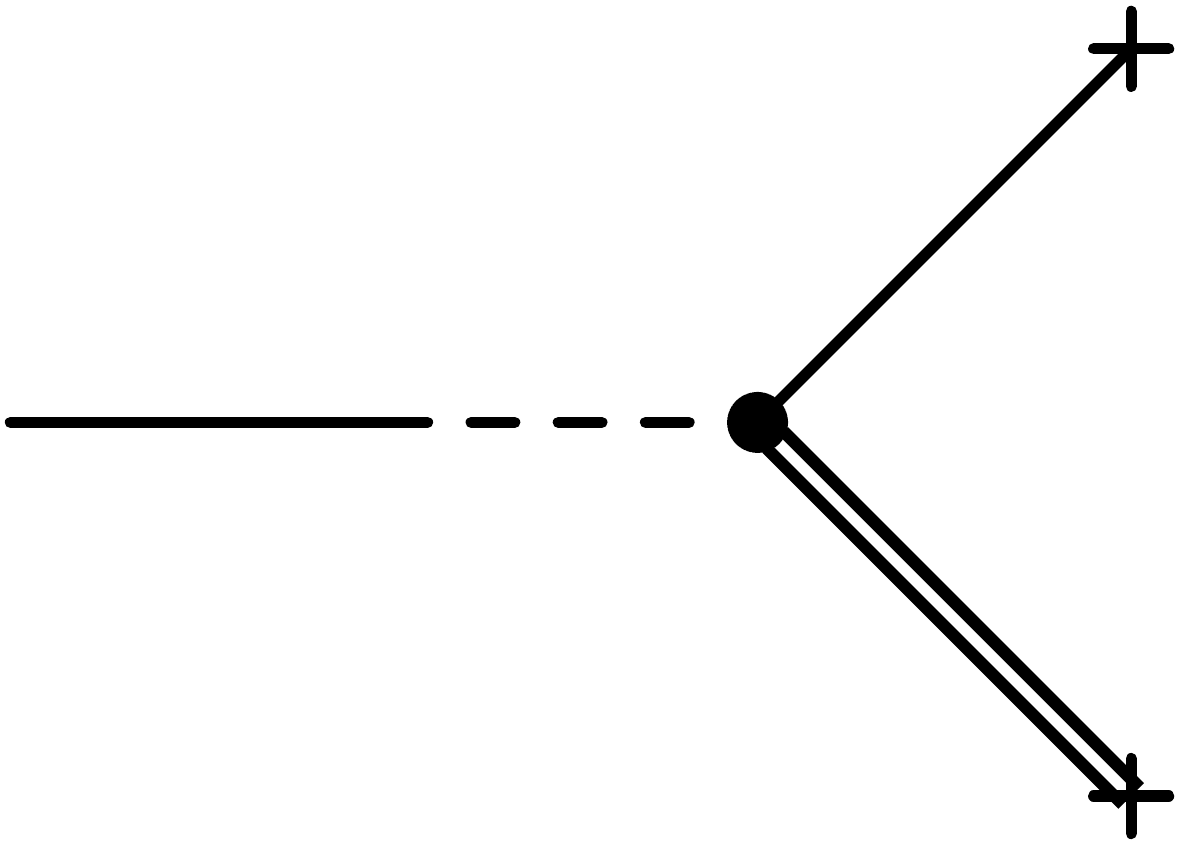} 
\hspace{0.4in}
\includegraphics[scale=0.2]{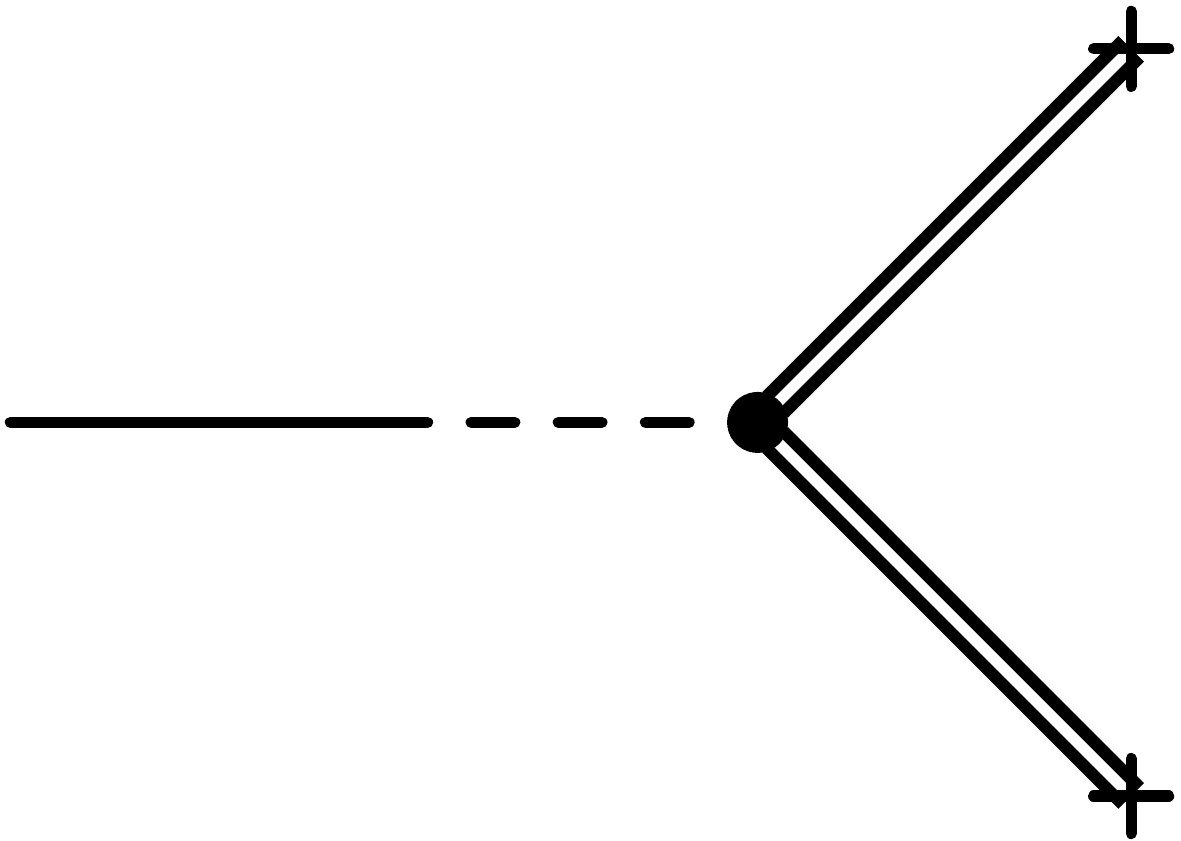}
\caption{Inflaton fluctuation corrections ${\bar \varphi}^{(1)}$ induced by interactions.  Crosses indicate the zeroth-order solution ${\bar \varphi}^{(0)}$ or~${\bar \chi}^{(0)}$.}
\label{classical_perturbations}
\end{center}
\end{figure}

We may then iteratively solve for higher-order solutions to the fluctuations,
\begin{equation}
\hspace{-0.1in} {\bar \varphi}^{(1)}(x) = \int d^4 y \sqrt{g} \ G^R (x,y) \left( g_1 {\bar \varphi}^{(0)} {\bar \chi}^{(0)}(y) + \frac{g_2}{2} {\bar \chi}^{(0)}(y)^2  \right).
 \end{equation}
Fourier transforming into comoving momentum, these vertices can be evaluated using the retarded Green's function $G^{R}$ can be written in terms of the fluctuation solutions,
\begin{eqnarray}
\label{keldyshgreens}
G^R_{\bf k}(\tau_1,\tau_2) &\equiv& i \langle {\bar \varphi}^{(0)}_{\bf k} (\tau_1) \Phi^{(0)}_{\bf -k}(\tau_2) \rangle \\
\nonumber
&=& - 2 \theta(\tau_1-\tau_2) {\rm Im} \left[ U_{\bf k}(\tau_1) U^*_{\bf k}( \tau_2) \right] .
\end{eqnarray}
These are shown in Figure~\ref{classical_perturbations}.  The advanced Green's function $G^A$ is then simply the time-reversal of this:
\[ G^A_{\bf k}(\tau_1,\tau_2) \equiv G^R_{\bf k}(\tau_2,\tau_1) . \]

Note that comoving momentum is conserved at vertices.  A similar procedure applies for ${\bar \chi}^{(1)}$ and ${\bar \gamma}^{(1)}$ using their corresponding retarded Green's function $\mathcal G^R$ and $\mathbb G^R$,
\begin{eqnarray}
\label{keldyshgreens}
\mathcal G^R_{\bf k}(\tau_1,\tau_2) &\equiv& i \langle {\bar \chi}^{(0)}_{\bf k} (\tau_1) {\rm X}^{(0)}_{\bf -k}(\tau_2) \rangle \\
\nonumber
&=& - 2 \theta(\tau_1-\tau_2) {\rm Im} \left[ V_{\bf k}(\tau_1) V^*_{\bf k}( \tau_2) \right] , \\
\nonumber 
\mathcal G^A_{\bf k}(\tau_1,\tau_2) &\equiv& \mathcal G^R_{\bf k}(\tau_2,\tau_1), \\
\mathbb G^R_{\bf k}(\tau_1,\tau_2) &\equiv& i \langle {\bar \gamma}^{(0)}_{\bf k} (\tau_1) \Gamma^{(0)}_{\bf -k}(\tau_2) \rangle \\
\nonumber
&=& - 2 \theta(\tau_1-\tau_2) {\rm Im} \left[ W_{\bf k}(\tau_1) W^*_{\bf k}( \tau_2) \right] , \\
\nonumber 
\mathbb G^A_{\bf k}(\tau_1,\tau_2) &\equiv& \mathbb G^R_{\bf k}(\tau_2,\tau_1). 
\end{eqnarray}

Of course the solution may then be iterated again to obtain yet even higher-order solutions,
\[ {\bar \varphi}^{(2)}(x) = \int d^4 y \sqrt{g} \ G^R (x,y) \left( g_1 {\bar \varphi}^{(0)} {\bar \chi}^{(1)}(y) + \cdots \right). \]
Note that for fluctuations the slow-roll parameters $\epsilon_i$ are implicitly included, and that the order of the solution refers to the order of the couplings $g_i$.

\subsubsection{Statistical Correlations}
\begin{figure}
\begin{center}
\hspace{0.4in}
\parbox{34mm}{\includegraphics[scale=0.16]{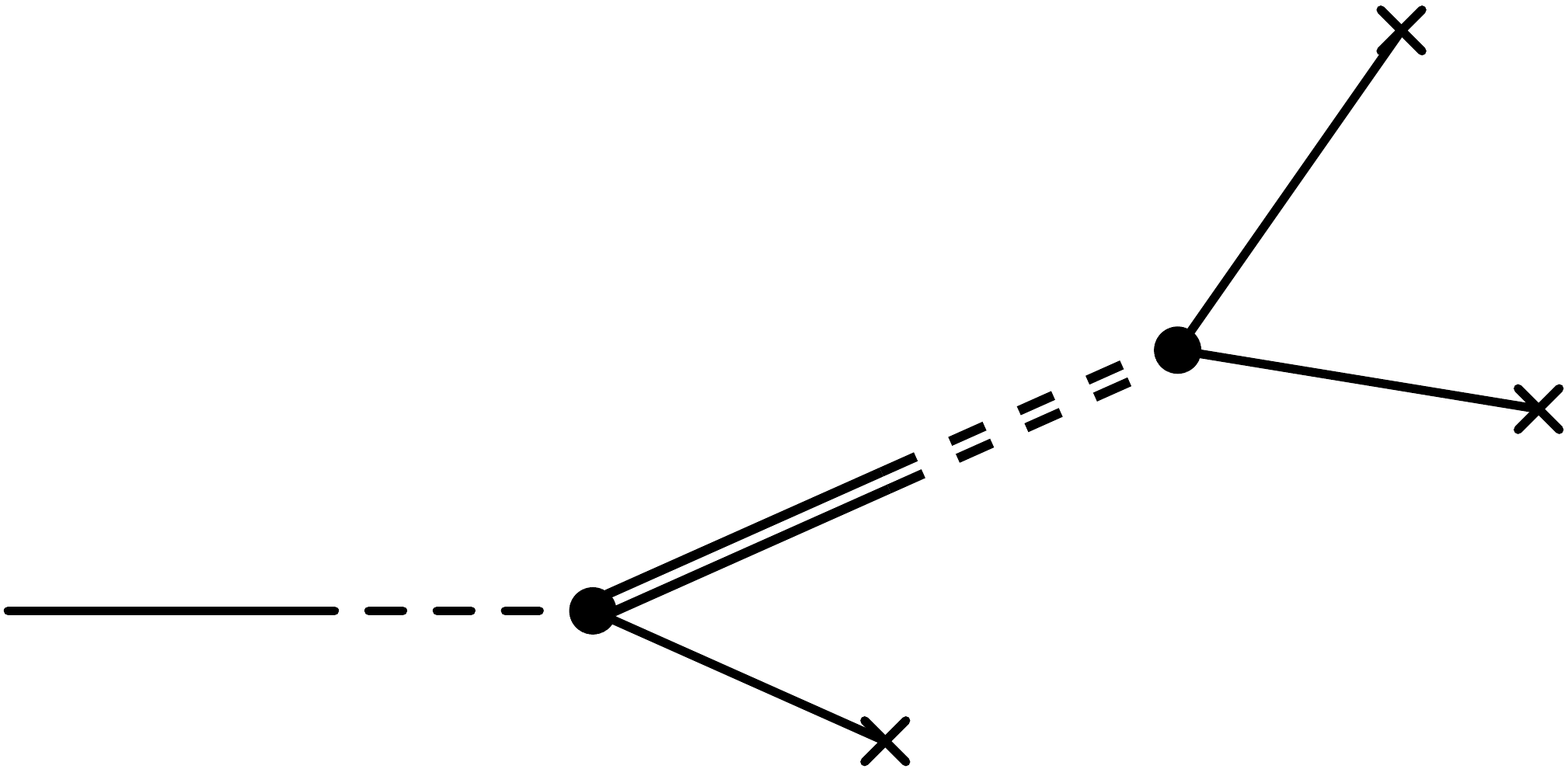}}
$\ \ \ \times$
\parbox{34mm}{\includegraphics[scale=0.16]{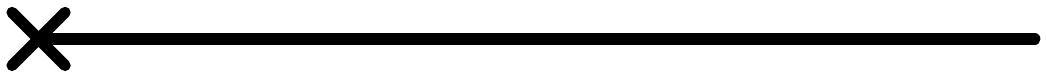}}
\newline
$\ \Longrightarrow \ $
\parbox{27mm}{\includegraphics[scale=0.16]{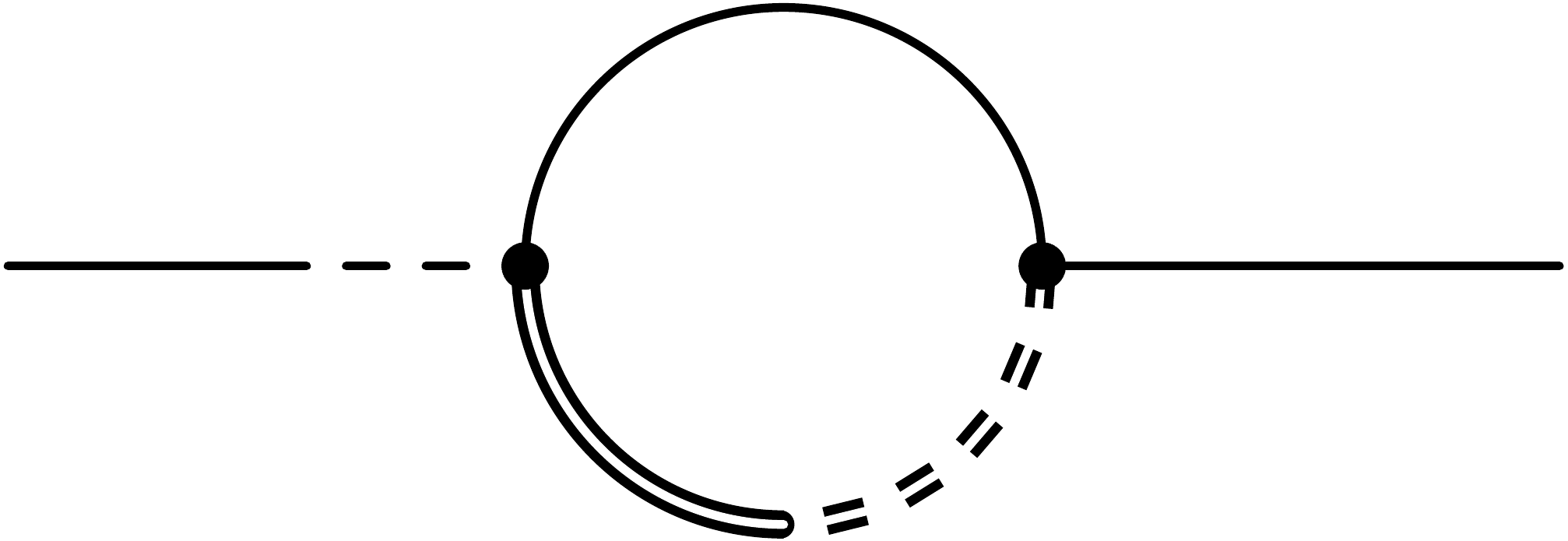}}
$\ \ \ \ \ \ \ +$
\parbox{27mm}{\includegraphics[scale=0.16]{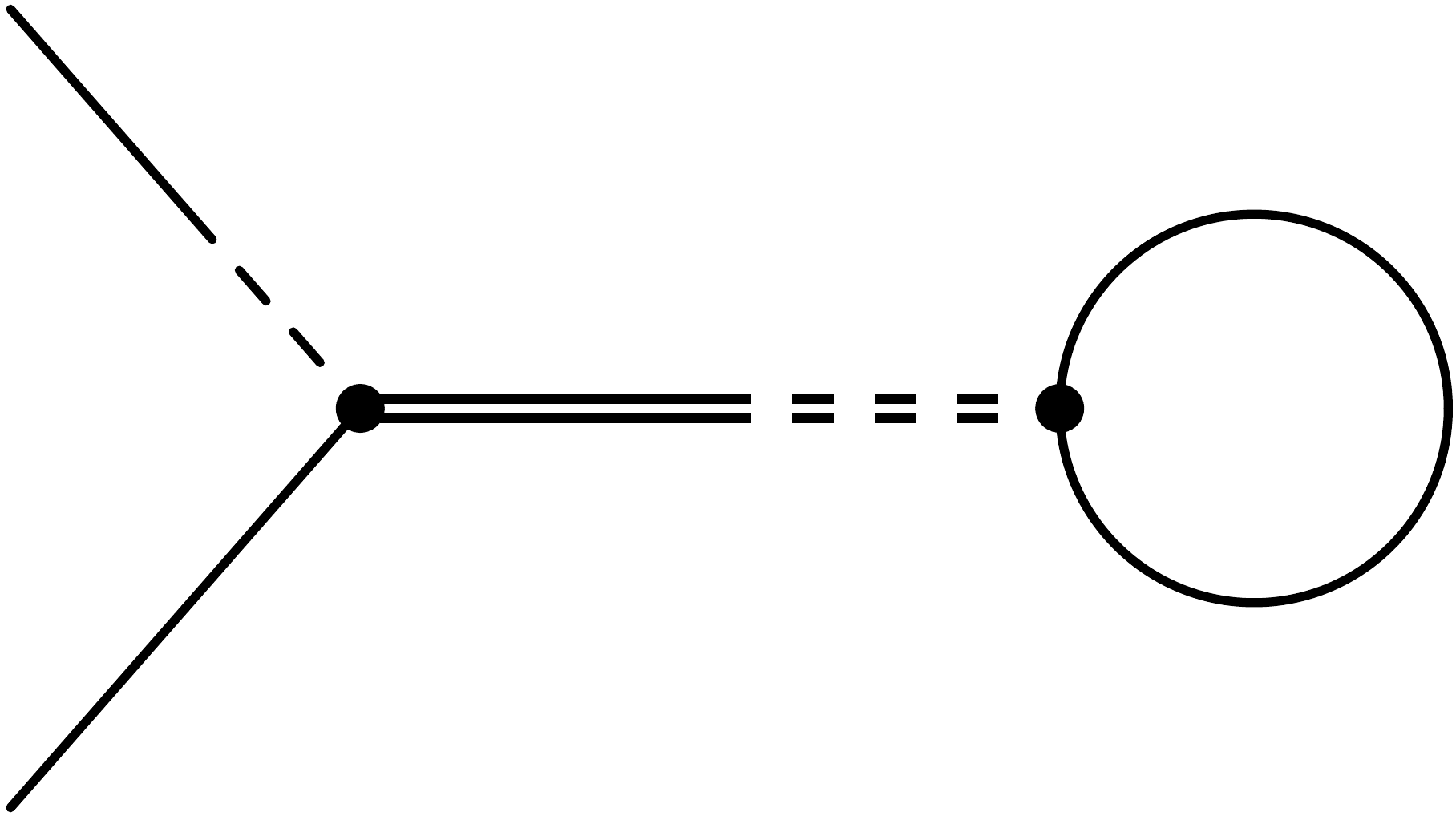}} 
\caption{Statistical averaging of a tree-level solution may produce loops, despite remaining classical.  Note that the second diagram can always be cancelled with an appropriate counterterm.}
\label{classical_corrections}
\end{center}
\end{figure}

After obtaining ${\bar \varphi}$ to our desired order, we may then take statistical averages of the zeroth order solutions via 
\begin{eqnarray*}
F_{\bf k}(\tau_1,\tau_2) &=& \langle {\bar \varphi}^{(0)}_{\bf k}(\tau_1) {\bar \varphi}^{(0)}_{\bf -k}(\tau_2) \rangle \\
&=& {\rm Re} \left[ U_{\bf k}(\tau_1) U^*_{\bf k}( \tau_2) \right] ,  \\
0 &=&\langle \Phi^{(0)}_{\bf k}(\tau_1) \Phi^{(0)}_{\bf -k}(\tau_2) \rangle, \\
\mathcal F_{\bf k}(\tau_1,\tau_2) &=& \langle {\bar \chi}^{(0)}_{\bf k}(\tau_1) {\bar \chi}^{(0)}_{\bf -k}(\tau_2) \rangle \\
&=& {\rm Re} \left[ V_{\bf k}(\tau_1) V^*_{\bf k}( \tau_2) \right] ,  \\
0 &=&\langle {\rm X}^{(0)}_{\bf k}(\tau_1) {\rm X}^{(0)}_{\bf -k}(\tau_2) \rangle, \\
\mathbb F_{\bf k}(\tau_1,\tau_2) &=& \langle {\bar \gamma}^{(0)}_{\bf k}(\tau_1) {\bar \gamma}^{(0)}_{\bf -k}(\tau_2) \rangle \\
&=& {\rm Re} \left[ W_{\bf k}(\tau_1) W^*_{\bf k}( \tau_2) \right] ,  \\
0 &=&\langle \Gamma^{(0)}_{\bf k}(\tau_1) \Gamma^{(0)}_{\bf -k}(\tau_2) \rangle
 \end{eqnarray*}
to get correlations.  This can be heuristically thought of as gluing the crosses together in all possible ways.  An example is shown in Figure~\ref{classical_corrections}.
While this may produce Feynman diagrams with loops, representing integrals over comoving momentum, \emph{they are nonetheless completely classical}.  An important difference of the in-in formalism is that loops may represent statistical, but not quantum, fluctuations. 

\subsection{Self-Consistency of Background Solution}
\begin{figure}
\begin{center}
\includegraphics[scale=0.2]{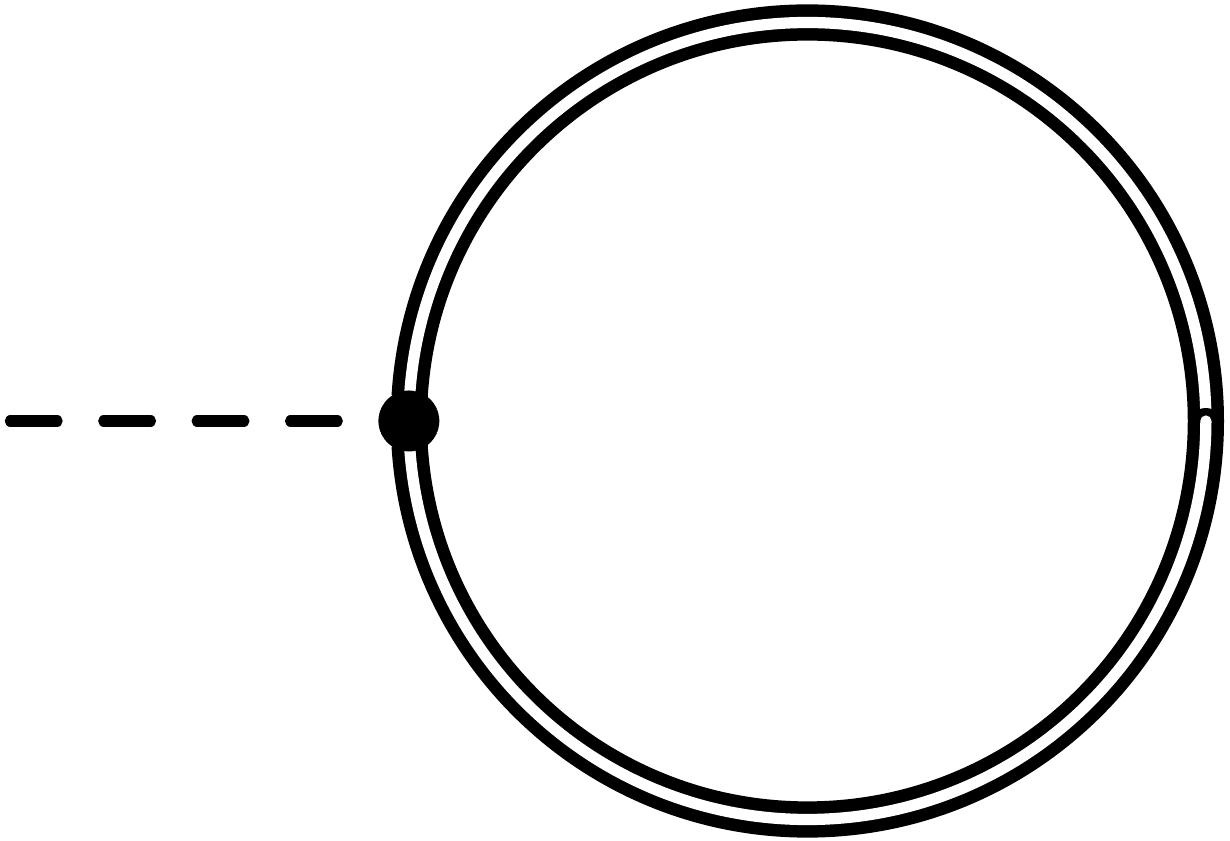} \\
\caption{Correction to the background linear coupling coming from $\chi$-fluctuation backreaction.  This can be absorbed into the definition of $j$.}
\label{linear_bg_corrections}
\end{center}
\end{figure}
The introduction of fluctuations means that the background solution will be slightly modified due to backreaction.  This is an $\mathcal O(g_2)$ correction to the linear coupling $j$, as shown in Figure~\ref{linear_bg_corrections}, having the value
\[\Delta j = \langle \int \frac{d^3 {\bf q}}{(2 \pi)^3} \frac{g_2}{2} {\bar \chi}^{(0)}_{\bf q}  {\bar \chi}^{(0)}_{\bf -q}(\tau) \rangle =   \frac{g_2}{2}  \int \frac{d^3 {\bf q}}{(2 \pi)^3} \mathcal F_{\bf q}(\tau,\tau) . \]
The apparent time-dependence is illusory, as follows.  Let us convert the integral over comoving momentum into one over physical momentum.  In static-background QFT loop regularization one imposes a UV cutoff $\Lambda$ on the Wick-rotated 4-momentum, thus respecting Lorentz symmetry.  Since the quasi-de Sitter background of inflation breaks Lorentz symmetry (`energy' is ill-defined) we cannot impose such a cutoff, but comoving momentum \emph{is} conserved, and therefore we can perform an equivalent procedure as follows.  Via (\ref{chisol}) the $\chi$-field's conjugate variable to time is given by 
\[ E \sim \sqrt{q^2 \tau^2 H^2 + M^2} .  \]
Thus we may place a cutoff on $E$ at some scale, effectively imposing a cutoff on $q$.  Let us momentarily assume that $H$ is constant, allowing us to estimate the correction as
\begin{eqnarray*}
\Delta j &\approx& \frac{g_2}{2} \int_{0}^{\sqrt{\Lambda^2-M^2}/H |\tau|} \frac{d^3 {\bf q}}{(2 \pi)^3} \frac{H^2  \tau^2}{2 \sqrt{ q^2 + \frac{M^2}{H^2 \tau^2}}} \\
&=& \frac{g_2}{2 \pi} \left[ \Lambda \sqrt{\Lambda^2-M^2} + M^2 \ln \left( \frac{M}{\Lambda + \sqrt{\Lambda^2-M^2} }  \right) \right] .
\end{eqnarray*}
This is independent of $H$, assuring us that it is the same answer one would get from a static spacetime answer had we simply truncated $E \leq \Lambda$.  This time-independent correction can then be cancelled by appropriate redefinition of $j$.  A similar procedure can of course be done for the tadpole of $\chi$, as shown in Figure 3.

\section{Power Spectrum Evaluation}
Here we will give the explicit formula for computing the scalar and tensor power spectrum in our high-energy model.  
\subsection{Inflaton Fluctuations}
\begin{figure}
\begin{center}
\parbox{40mm}{\includegraphics[scale=0.18]{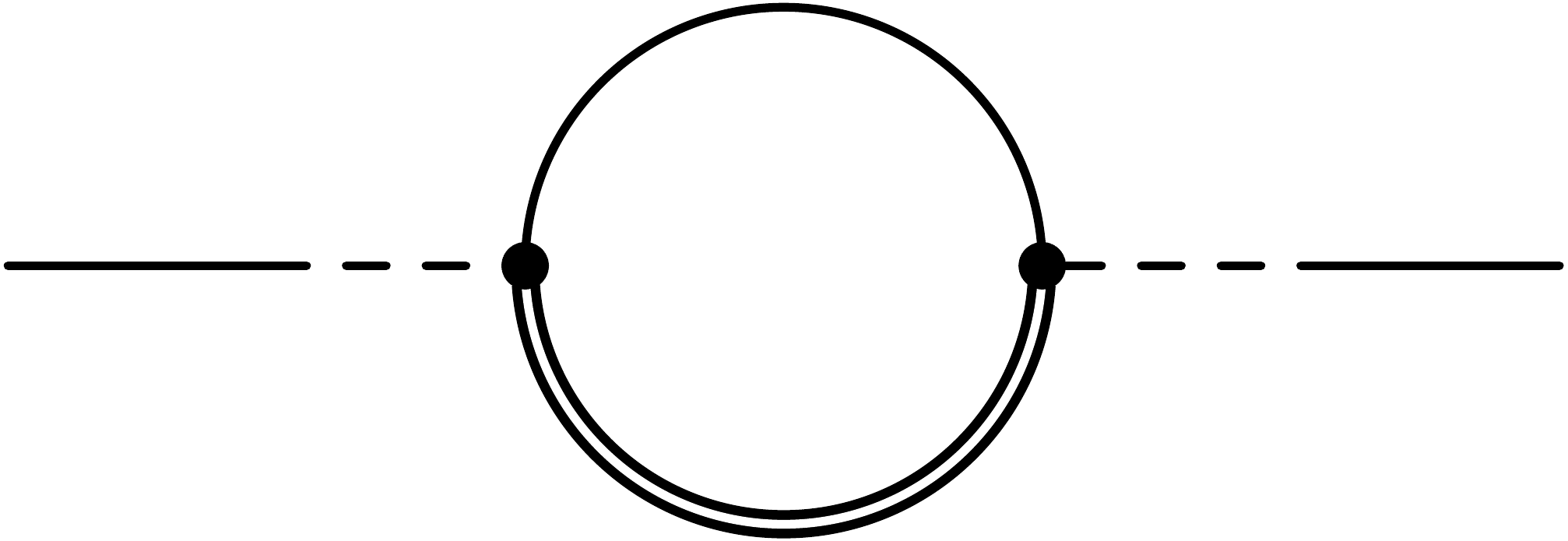} \hspace{-0.1in} \newline A}
\parbox{40mm}{\includegraphics[scale=0.18]{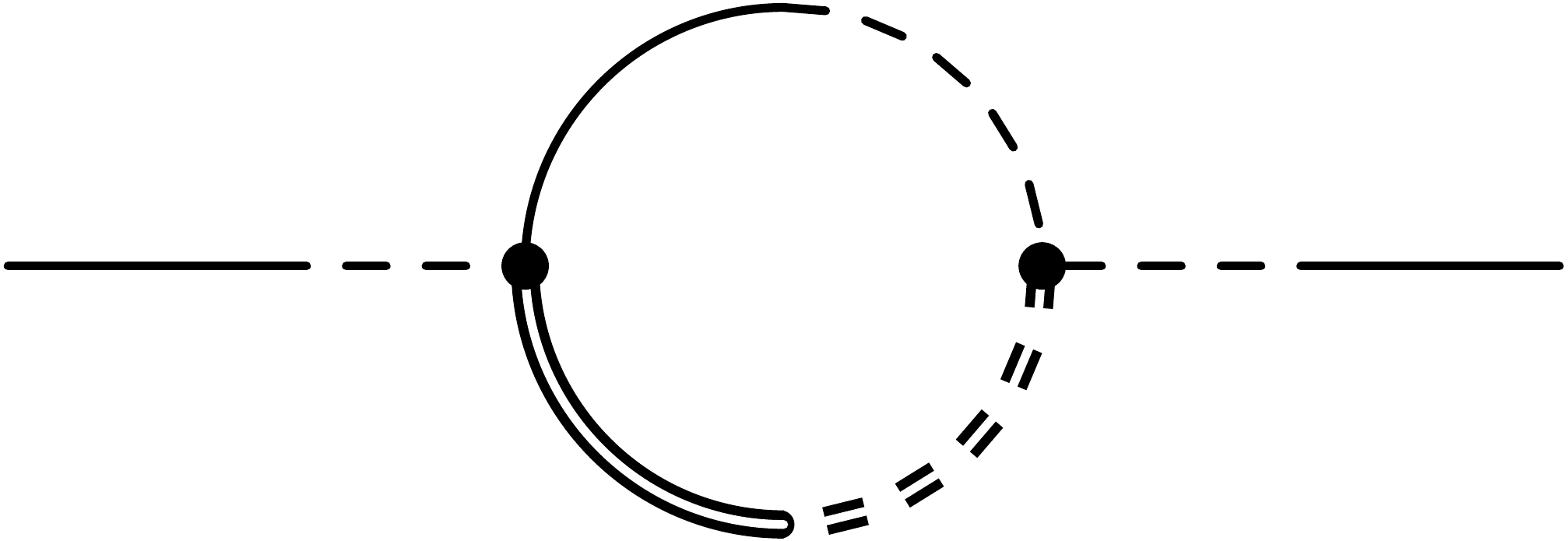} \hspace{-0.1in} \newline B} \\
\vspace{0.1in}
\parbox{40mm}{\includegraphics[scale=0.18]{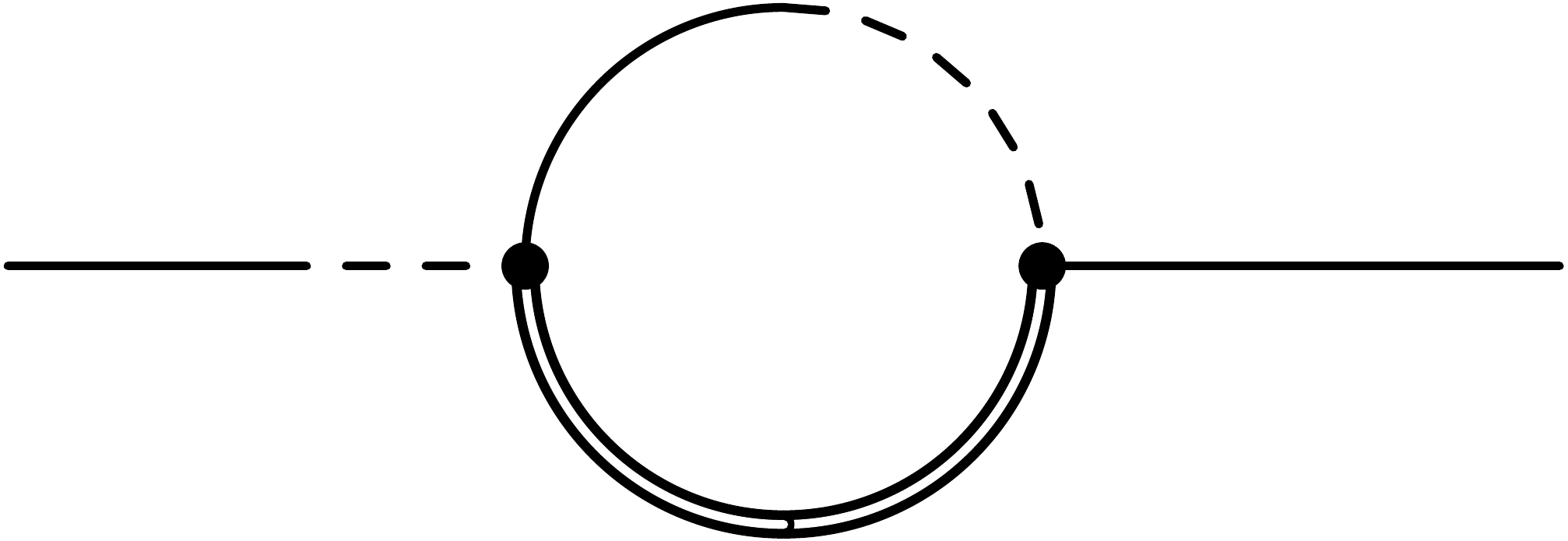} \hspace{-0.1in} \newline C} 
\parbox{40mm}{\includegraphics[scale=0.18]{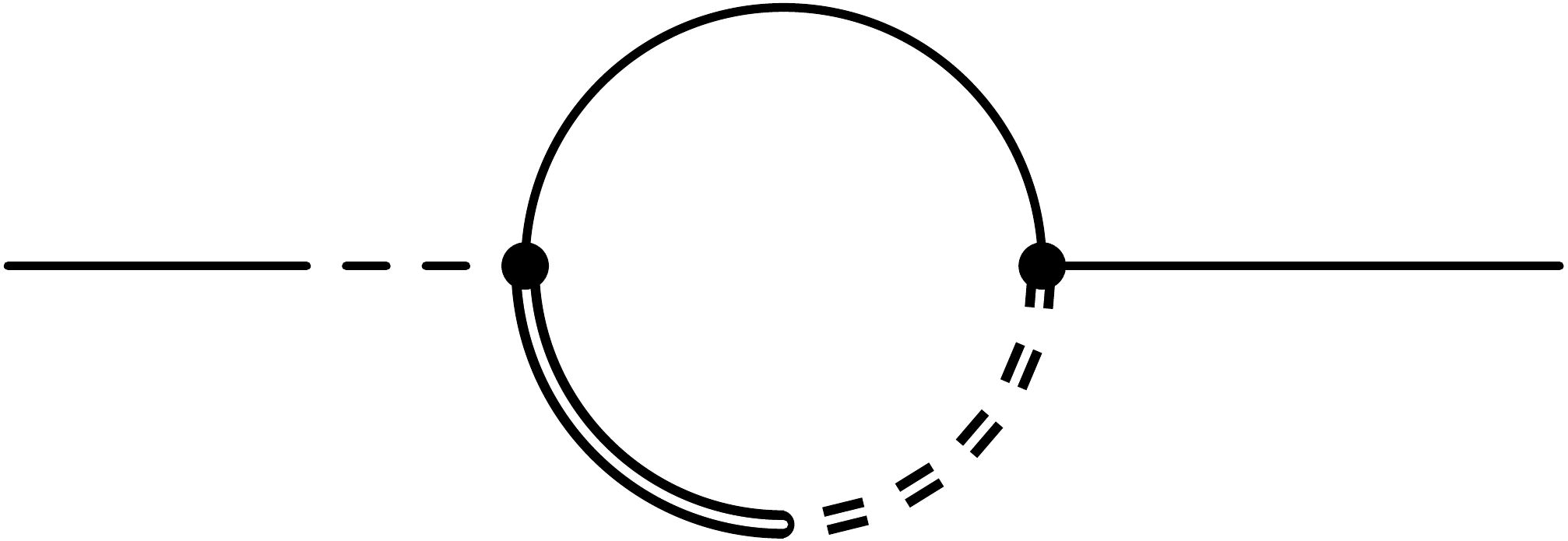} \hspace{-0.1in} \newline D} 
\caption{Power spectrum corrections mediated by the heavy field.  Single solid lines indicate contractions of ${\bar \varphi}$, dashed single lines indicate those of $\Phi$, with analogous notation for double lines indicating the heavy field components $\{ {\bar \chi},{\rm X} \}$.}
\label{ps_corrections}
\end{center}
\end{figure}

We will need to know the correlation of inflaton fluctuations at the moment of horizon crossing,
\[ P_{\varphi}(k) \equiv \frac{k^3}{2 \pi^2} \left. \langle  {\bar \varphi}_{\bf k} (\tau)  {\bar \varphi}_{\bf -k} (\tau) \rangle \right| _{k = aH}.  \]
To evaluate this perturbatively, substitute the classical solution ${\bar \varphi} = {\bar \varphi}^{(0)} +  {\bar \varphi}^{(1)} + \cdots$ into this expression.  In the decoupling limit $g_i \rightarrow 0$ or $M \rightarrow \infty$, the inflaton fluctuation power spectrum is simply
\begin{eqnarray*}
\label{ps0}
P^{(0)}_{\varphi}(k) &=& \frac{k^3}{2 \pi^2} \langle  {\bar \varphi}^{(0)}_{\bf k}(\tau_k)  {\bar \varphi}^{(0)}_{\bf -k}(\tau_k) \rangle \\
&=& \frac{k^3}{2 \pi^2} F_{\bf k}(\tau_k,\tau_k) 
\end{eqnarray*}
where $\tau_k$ is the horizon-crossing time of mode $k$, $\tau_k~\sim~-~1/k$.  The first order corrections are then:
\[ P^{({\rm 1})}_{\varphi}(k) = \frac{k^3}{2 \pi^2} \times 2 \langle {\bar \varphi}_{\bf k}^{(1)} {\bar \varphi}_{\bf -k}^{(0)} \rangle . \]
There are no interactions in the action (\ref{keldyshinteractions}) which will produce this, and so we turn to the second order contributions,
\begin{equation}
\label{order2corrects}
 P^{({\rm 2})}_{\varphi}(k) = \frac{k^3}{2 \pi^2} \left( \langle {\bar \varphi}_{\bf k}^{(1)} {\bar \varphi}_{\bf -k}^{(1)} \rangle + 2 \langle {\bar \varphi}_{\bf k}^{(2)} {\bar \varphi}_{\bf -k}^{(0)} \rangle  \right).
 \end{equation}
There are a total of four diagrams which could possibly contribute at $\mathcal O(g_1^2)$, shown in Figure~\ref{ps_corrections}.  We will give explicit expressions for the first two,
\begin{eqnarray*}
 P^{({\rm A})}_{\varphi}(k) &=& \frac{k^3}{2 \pi^2} \left( -i g_1 \right)^2  \int_{\tau_{\rm in}}^{\tau_k} d \tau_1 \ a(\tau_1)^4 \int_{\tau_{\rm in}}^{\tau_k} d \tau_2 \ a(\tau_2)^4 \times \\
&& \hspace{-0.9in} \int \frac{d ^3 \bf q}{(2 \pi)^3} \left[ -i G^R_{\bf k}(\tau_k,\tau_1) \right]  \mathcal F_{\bf q+k}(\tau_1,\tau_2) F_{\bf q}(\tau_1,\tau_2) \left[ -i G^A_{\bf k}(\tau_2,\tau_k) \right], \\
 P^{({\rm B})}_{\varphi}(k) &=& \frac{k^3}{2 \pi^2} \left( -i g_1 \right)^2  \int_{\tau_{\rm in}}^{\tau_k} d \tau_1 \ a(\tau_1)^4 \int_{\tau_{\rm in}}^{\tau_k} d \tau_2 \ a(\tau_2)^4 \times \\
&& \hspace{-0.9in}  \int \frac{d ^3 \bf q}{(2 \pi)^3} \left[ -i G^R_{\bf k}(\tau_k,\tau_1) \right] [ - i G^R_{\bf k}(\tau_1,\tau_2)] \left[ -i \mathcal G^R_{\bf 0}(\tau_1,\tau_2) \right] [-i G^A_{\bf q}(\tau_2,\tau_k) ].
 \end{eqnarray*}

\subsection{Inflaton Vertex Evaluation}
We first dissect the interaction at the vertex before evaluating the full diagram.  Since the time coordinate is integrated over, vertex evaluation is very different than in standard quantum field theory.  Writing out the Green's and Wightman's functions in terms of $U$'s and $V$'s, we see there are three types of cubic vertices.  We will evaluate each.

The first is
\begin{equation}
 \label{vertexA}
 \mathcal A_1( \tau_k; {\bf k}_1,{\bf k}_2) \equiv \int _{\tau_{\rm in}}^{\tau_k} d \tau \ a(\tau)^4 {U}_{{\bf k}_1}(\tau)  {U}_{{\bf k}_2}(\tau)  V^*_{-({\bf k}_1+{\bf k}_2)}(\tau).
 \end{equation}
The superscript $k$ indicates there is some dependence upon the moment that the mode leaves the horizon.  Since we assume all interactions happen well inside the horizon, a good approximation is
\[ U_{\bf k}(\tau) \approx \frac{1}{a(\tau) \sqrt{2 k}} \exp \left[ -i \left( k \tau + \frac{\nu \pi}{2} + \frac{\pi}{4} \right) \right], \hspace{0.2in} k | \tau| \gg 1  \]
where $\nu$ is given by (\ref{varphisol}).  To leading order in the slow-roll parameters, the scale factor is given by
\[ a(\tau) \approx - \frac{1+\epsilon_1}{ \tau H} . \]
Substituting this into (\ref{vertexA}) results in the expression
\begin{equation}
 \mathcal A_1( \tau_k; {\bf k}_1,{\bf k}_2) \approx - \frac{1}{2  \sqrt{2k_1 k_2} } \int _{\tau_{\rm in}}^{\tau_k} \frac{d \tau}{ H \tau}  \frac{ (1+\epsilon_1) e^{-i \phi} }{\left( | {{\bf k}_1+{\bf k}_2}|^2 + \frac{M^2}{H^2 \tau^2} \right)^{1/4} } 
\end{equation}
where the phase is given by
\[ \phi(\tau) \equiv \nu \pi + (k_1 + k_2) \tau - \int ^{\tau}_{\tau_{\rm in}} d \tau' \sqrt{ | {{\bf k}_1+{\bf k}_2}|^2 + \frac{M^2}{H^2 \tau'^2}} . \]
Let us now introduce the rescaled time coordinate $u$ such that
\[ u \equiv \frac{H}{M} \tau. \]
To obtain the differential, recall that the Hubble scale changes with time as 
\[ H(\tau) \approx H_* \left( \frac{\tau}{\tau_*} \right)^{\epsilon_1} \]
where $\tau_*$ is some reference time and $H_*$ is some the Hubble scale at this time.  Then
\[ du = \frac{H}{M} \left( 1 + \epsilon_1 \right) d \tau. \]
In this variable it is clear that $\mathcal A_1$ admits a stationary phase approximation at the energy-conservation moment
\begin{eqnarray*}
0 = \frac{d\phi}{du} &=& \frac{M/H}{1 + \epsilon_1} \frac{d \phi}{d \tau} \\
&=& \frac{M/H}{1 + \epsilon_1} \left[  \pi \frac{d \nu}{d \tau} + k_1+ k_2 - \sqrt{  | {\bf k}_1+{\bf k}_2|^2 +u^{-2}_c}\right]  . 
\end{eqnarray*}
The variation in slow-roll phase is given by
\[ \frac{d \nu}{d \tau} = \frac{dN}{d \tau} \frac{d \nu}{dN} \approx - \frac{3}{\tau} \epsilon_1 \epsilon_2 \]
and so is higher-order in slow-roll and can be neglected.  The solution then defines the NPH,
\[ u_{c}^{-1}= - \sqrt{2k_1k_2(1-\cos\theta)}, \hspace{0.3in} \cos\theta=\frac{ {\bf k}_1 \cdot {\bf k}_2}{k_1 k_2}. \]
Expanding the phase near this stationary point gives the inflection
\[ \left. \frac{d^2 \phi}{du^2} \right|_{ u_c} =  \left. \frac{M/H}{1 + \epsilon_1} \right|_{u_c}  \frac{u_c^{-3}}{k_1+k_2} . \]
Then to leading order in $H/M$ and slow-roll parameters the amplitude is
\begin{eqnarray}
{\mathcal A}_1(\tau_k; {\bf k}_1,{\bf k}_2)  &\approx&\\
\nonumber
&& \hspace{-0.8in} - \left. \frac{ \sqrt{\pi i}\theta(\tau_k-\tau_c) \theta(\tau_c - \tau_{\rm in})(1+ \frac{3}{2} \epsilon_1)e^{-i \phi}}{ 2\sqrt{k_1 k_2} \left[ 2 k_1 k_2 (1- \cos \theta ) \right]^{1/4} \sqrt{HM}}\right|_{\tau_c}  .
\end{eqnarray}
The physics of this is clear. This diagram accounts for the threshold production/decay of heavy particles at high redshift in the early universe.  Although this induced term is localized in time, it is $k$-dependent and thus does not represent a physical boundary.  Rather, it represents a `hypersurface' representing the energy scale of interaction, precisely as expected in the New Physics Hypersurace approach. 

The second possible vertex is identical to $\mathcal A_1$ but with one $U$ conjugated,
\[ \mathcal A_2(\tau_k; {\bf k}_1,{\bf k}_2) \equiv \int _{\tau_{\rm in}}^{\tau_k} d \tau \ a(\tau)^4 {U}_{{\bf k}_1}(\tau)  {U}^*_{{\bf k}_2}(\tau)  V^*_{-({\bf k}_1+{\bf k}_2)}(\tau). \]
This has only imaginary-time saddlepoint solutions.  Since our $\tau$-integral is confined to the real axis we will never pass over this point in our integration, and so this amplitude will be suppressed as $\mathcal A_2 \sim {\rm erf} (\frac{M}{H}) \sim \frac{H}{M} e^{-(M/H)^2}$ except for $\cos \theta \approx -1$, when the saddlepoint moves to infinity and the inflection vanishes.  In all other cases we may neglect such interactions.  Finally we consider $\mathcal A_3$ which has both $U$'s conjugated and so admits no saddlepoint solutions, and thus can also be neglected.

\subsection{Inflaton Power Spectrum}
\begin{figure}
\begin{center}
\includegraphics[width=3in]{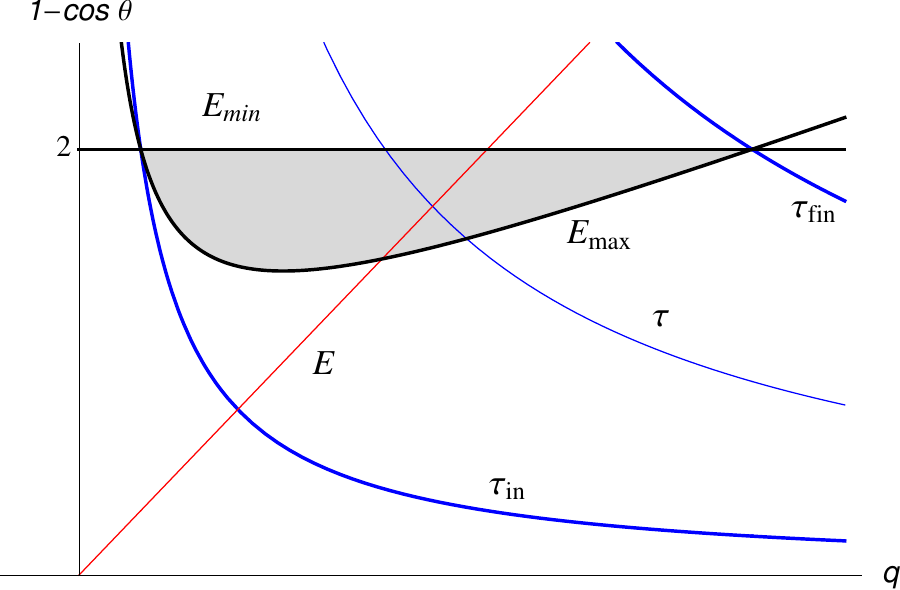}
\caption{The phase space of diagram $A$ in both $\{ q, 1- \cos \theta \}$ and $\{ \tau,E \}$ coordinate systems.  Constant-$\tau$ contours are shown in blue, and a constant-$E$ contour is shown in red.  The low-energy phase space allowed in our integration is shown in gray.}
\label{loopspace}
\end{center}
\end{figure}
After the vertex evaluation it remains to perform the integration over loop momentum.  
\subsubsection{First scalar correction}
Let us first consider diagram $A$,
\begin{equation}
\label{pa}
 P^{({\rm A})}_{\varphi}(k) = \frac{g_1^2}{2} \frac{k^3}{2 \pi^2} \left| U_{\bf k}(\tau_k) \right|^2 \int \frac{d ^3 \bf q}{(2 \pi)^3} \left| \mathcal A_1( \tau_k; {\bf k}, {\bf q} ) \right|^2  . 
 \end{equation}
This is non-trivial for two reasons.  First, we wish to implement a physical cutoff, and so a change of coordinates is required.  Second, loop momentum influences the time of vertex interaction via $\tau_c( {\bf q}, {\bf k})$, and so affects the Hubble scale.  Thus we need to convert the integral over comoving momentum to one over time of interaction.  To accomplish this, first simplify the 3d integral to 2d,
\[ \int \frac{ d^3 {\bf q}}{(2 \pi)^3} \rightarrow \frac{1}{(2 \pi)^2} \int q^2 dq d(1-\cos \theta) .\]
Now replace the coordinates $\{ q,1-\cos \theta \}$ with $\{ \tau, E \}$ via the definitions 
\begin{eqnarray*}
\tau^{-1} &\equiv& - \frac{H}{M} \sqrt{2 kq(1-\cos \theta)}, \\
E \equiv H q |\tau| &=& M \sqrt{ \frac{q}{2k(1-\cos \theta)} } . 
\end{eqnarray*}
Of course the definition of $\tau$ coincides with $\tau_c$, since this is the stationary-phase localized time of interaction.  Inverting these coordinate systems gives
\begin{equation}
\label{tausigma}
q =  \frac{E}{|\tau| H}, \hspace{0.4in} 1- \cos \theta = - \frac{M^2}{2 H E \tau  k}  . 
\end{equation}
The measure transforms as
\[ d (1 - \cos \theta) dq \rightarrow \frac{ M^2(1+ \epsilon_1)}{H^2 E \tau^3 k} dE d \tau. \]
The integral in (\ref{pa}) is then transformed as
\begin{eqnarray*}  
&& \int \frac{d ^3 \bf q}{(2 \pi)^3} \left| \mathcal A_1(\tau_k; {\bf k}, {\bf q} ) \right|^2  \\
&=& \frac{1}{16 \sqrt{2} \pi k^{3/2} M} \int \frac{ \sqrt{q} dq d (1-\cos \theta)}{H \sqrt{1-\cos \theta}} (1 + 3 \epsilon_1) \\
&\rightarrow& - \frac{1}{16 \pi k^2 } \int \frac{d\tau d E}{\tau^{3} H^4} (1+4 \epsilon_1) . 
\end{eqnarray*}
The phase $\phi$ cancels out in this amplitude and so we need not evaluate it.  We can now easily place limits on the region of integration in terms of the energy scale $\Lambda$:

\paragraph{Maximal $Q$} As done previously, we place a limit on the physical energy to be $\Lambda$.  Using the fact that the vertex with a virtual $\varphi$ and an external $\varphi$ of comoving momentum ${\bf k}$ have precisely the energy of the $\chi$-field, the bound is found to be
\[ E_{\rm max} = - H k |\tau| + \Lambda.  \]
\paragraph{Minimal $E$} Since $1- \cos \theta \leq 2$ this places a minimal limit on $E$,
\[ E_{\rm min} = \frac{M^2}{4 H(\tau)^3 k |\tau|}. \]
Although this is time-dependent, it remains scale-invariant because it contains only the combination $k \tau$.
\paragraph{Initial and Final $\tau$} Of course we require $E_{\rm min}~<~E_{\rm max}$.  Their equality will determine the earliest and latest times for which this condition is valid, an equation which is quadratic in $\tau$.  The two solutions are given by 
\[ \tau_\pm = \tau^{(0)}_\pm \left( 1 - \epsilon_1 \ln \frac{ \tau^{(0)}_\pm}{\tau_*} \right), \]
where we have defined
\begin{equation}
\tau^{(0)}_\pm \equiv - \frac{\Lambda \pm \sqrt{\Lambda^2-M^2}}{2kH_*} .
 \end{equation}
 The later root is
\[ \tau_{\rm fin} =  \tau^{(0)}_- \left( 1 - \epsilon_1 \ln \frac{ \tau^{(0)}_-}{\tau_*} \right) , \]
whereas the earlier root is
 \[ \tau_{\rm in} =   \tau^{(0)}_+ \left( 1 - \epsilon_1 \ln \frac{ \tau^{(0)}_+}{\tau_*} \right). \]
This will always be after the point where the external energy parameter $E_k \equiv H k |\tau|$ reaches $\Lambda$, so we are assured that all energy scales under consideration are below $\Lambda$.  Figure~\ref{loopspace} shows the phase space in the two coordinate systems.  
 
This gives us a window $\tau_{\rm in} < \tau < \tau_{\rm fin}$ in which to integrate out this high-energy interaction in a controlled way.  Note that choosing $\Lambda \leq M$ makes this window vanish, a reassuring fact since (by definition)  there should not be any interactions below the scale of New Physics.  There is an upper bound on $\Lambda$ as well: the interactions must cease before the gauge-transformation at the horizon-exit near $\tau_k = -1/k$ (whereupon our plane-wave approximation would break down).  This means we are allowed to set the cutoff scale in the window
\[ M < \Lambda < \frac{1}{2} \left( H + M^2/H \right) . \]
For a typical value of $M/H \sim 100$, this means we can set $M \leq \Lambda \leq 50 M + H$, which should be adequate for the low-energy behavior.  We now expand the Hubble rate around the reference point $\ln \tau_* = - \ln k_*$, and keeping only the leading term in $\Lambda/M$, the integral is evaluated as
\begin{eqnarray*} 
&& \int_{ \tau_{\rm in} }^{ \tau_{\rm fin} } \frac{d \tau}{ \tau^{3} } \left(1 + 4 \epsilon_1 - 3 \epsilon_1 \ln \frac{\tau}{\tau_*}  \right) \int^{E_{\rm max}}_{E_{\rm min}} dE \\
& \approx& \left( - \frac{8 \Lambda^3 H_*^2 k^2}{3M^4} \right) \left[ 1 - \frac{2}{3} \epsilon_1 + 4 \epsilon_1 \ln \left| \frac{M^2}{4 \Lambda H_* k \tau_* } \right| \right] .
\end{eqnarray*}
The power spectrum correction is then
\begin{eqnarray}
\nonumber
P^{({\rm A})}_{\varphi}(k) &=&  \frac{g_1^2}{2} \frac{k^3}{24 \pi^3} \left| U_{\bf k}(\tau_k) \right|^2 \frac{\Lambda^3}{M^4 H_*} \\
\label{psa}
&\times& \left[ 1- \frac{2}{3} \epsilon_1 - 4 \epsilon_1 \ln \left( \frac{4 \Lambda H_* k}{ M^2 k_* } \right) \right] .
\end{eqnarray}

\subsubsection{Second scalar correction}
Diagram B can be evaluated in an identical fashion, except for two subtleties: the Heaviside function in the Green's function allow only half of the Gaussian fluctuations to contribute, and the two $i$'s produce a sign change.  Thus 
\[ P^{({\rm B})}_{\varphi}(k) = - \frac{1}{2} P^{({\rm A})}_{\varphi}(k). \]
\subsubsection{Third and fourth scalar correction}
Diagrams C and D can also be evaluated in a similar fashion, and to leading order in $H/M$ they cancel.
\subsubsection{Total result}
\begin{figure}
\begin{center}
\includegraphics[width=3in]{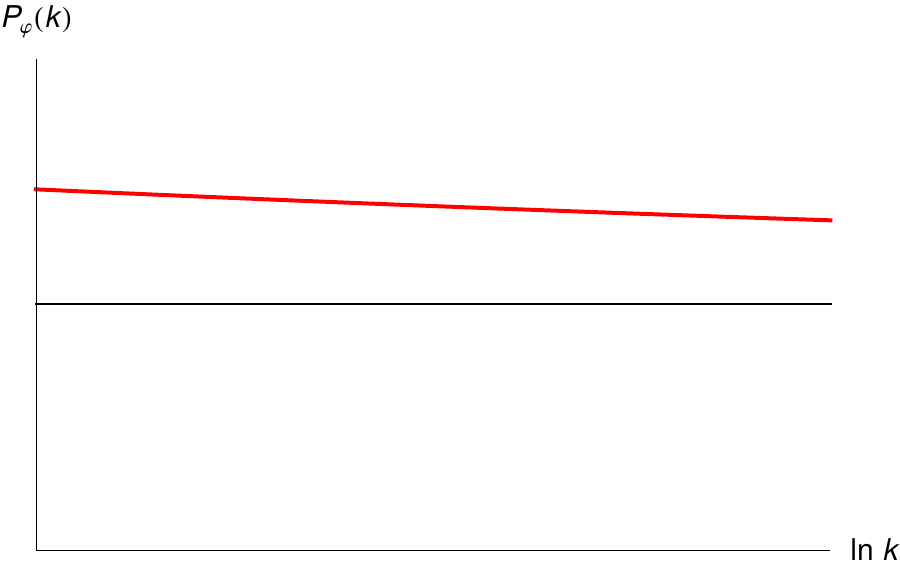}
\caption{Power spectra for the example theory in slow-roll inflation.  The red line shows the effect of interactions relative to that of the flat spectrum of the free theory in exact de Sitter space shown in black.  Note the corrections slightly decreases at higher $k$ as a result of the Hubble scale changing.}
\label{pk}
\end{center}
\end{figure}

Summing diagrams A, B, C and D results in the following total shift in the scalar power spectrum,
\begin{eqnarray}
\nonumber
\Delta P_{\varphi}(k) &=&  \frac{g_1^2 k^3}{96 \pi^3} \left| U_{\bf k}(\tau_k) \right|^2 \frac{\Lambda^3}{M^4 H_*} \\
\label{psa}
&\times& \left[ 1- \frac{2}{3} \epsilon_1 - 4 \epsilon_1 \ln \left( \frac{4 \Lambda H_* k}{ M^2 k_* } \right) \right] .
\end{eqnarray}
Since $\Lambda \sim M$ and $g_1 \sim H$, the correction then scales as $\Delta P \sim H/M$.  Figure~\ref{pk} shows these corrections to the power spectrum in the example theory. 

\subsection{The Tensor Power Spectrum}
We did not include a ${\gamma}^2 \chi$ coupling in our high-energy model, nor are there any induced by gauge-fixing; hence the tensor spectrum is the same as the standard slow-roll case.  The gravitational power spectrum is
\begin{eqnarray}
\nonumber
 P_{h,ij}^{(0)}(k) &\equiv& \frac{32k^3}{\pi M^2_{\rm pl}} \langle {\bar \gamma}^{(0)}_{ij,{\bf k}}(\tau_k) {\bar \gamma}^{(0)}_{ij,{\bf -k}} (\tau_k)\rangle \\
 \label{tensorps}
 &=&  \frac{ 32k^3}{\pi M^2_{\rm pl}}  \epsilon_{ij} \mathbb F(\tau_k,\tau_k).
 \end{eqnarray} 

\subsection{Comments and Interpretation}
\begin{figure}
\begin{center}
\includegraphics[scale=0.9]{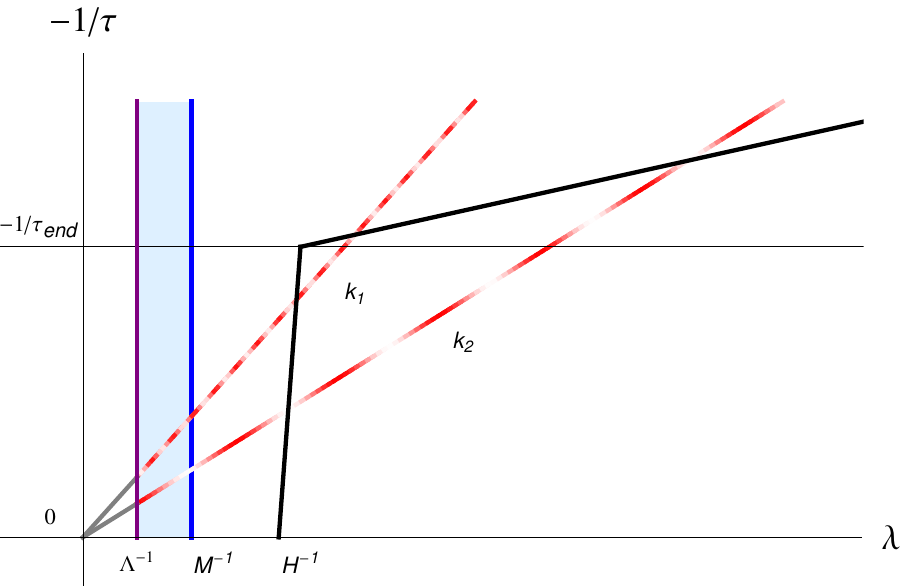}
\caption{The evolution of fluctuations during inflation, as they encounter the NPH.  The scale of New Physics $M$ is again indicated, as well as the cutoff scale $\Lambda$.  The comoving momentum of a fluctuation is given by its slope.  The high-energy interactions in the blue region can be effectively described as excited initial states for the inflaton.  The fact that all fluctuations encounter New Physics at (nearly) the same energy scale ensures that the answer is (nearly) scale-invariant.}
\label{TPschematic2}
\end{center}
\end{figure}

In the de Sitter limit of $\epsilon_i \rightarrow 0$, the result is a perfectly scale-invariant correction,
\[ \Delta P_\varphi = \frac{g_1^2 \Lambda^3 H_*}{192 \pi^3 M^4} . \]
Let us verify that we should have anticipated this by simply computing the physical momentum $p$ at the NPH for some fluctuation of comoving momentum $k$:
\[ p(\tau_c) = k/a(\tau_c) = - H(\tau_c) k \tau_c = M/2 . \]
So the NPH is indeed at a fixed energy scale for all modes.  A second verification is to compute the period of interaction near this NPH.  After all, we used a stationary phase approximation which includes the gaussian fluctuations around the stationary phase, giving the NPH a finite width.  Re-expressed in terms of physical time $\Delta t$ shows this interaction period also remains scale-invariant:
\[ \Delta t  = \frac{ d t}{d \tau} \Delta \tau \sim \left( -\frac{1}{H \tau_k} \right) \left( \sqrt{ \frac{ M}{H}} \frac{1}{k} \right) \sim \frac{1}{\sqrt{HM} }. \]
Figure~\ref{TPschematic2} shows a schematic of this.

The dependence upon $H/M$ is also worth commenting upon.  Recall the argument presented by \cite{Niemeyer:2000eh,Kempf:2000ac,Niemeyer:2001qe,Kempf:2001fa,Martin:2000xs,Brandenberger:2000wr,Brandenberger:2002hs,Martin:2003kp,Easther:2001fi,Easther:2001fz,Easther:2002xe,Kaloper:2002uj,Kaloper:2002cs,Danielsson:2002kx,Danielsson:2002qh,Shankaranarayanan:2002ax,Hassan:2002qk,Goldstein:2002fc,Bozza:2003pr,Alberghi:2003am,Schalm:2004qk,porrati,Porrati:2004dm,Hamann:2008yx,Achucarro:2010da,Starobinsky:2002rp,Chen:2010ef} in considering the low-energy expansions of local operators,
\[ S_{\rm bulk} = \int d^4 x \left[ \frac{1}{2} (\partial \varphi)^2 + \frac{\lambda}{M^2} \varphi \partial^4 \varphi + \cdots \right] . \]
Fluctuations frozen at $p \sim H$ will then have a power spectrum scaling as
\[ \langle \varphi^2 \rangle \sim H^2 \left( 1 + \frac{\lambda H^2}{M^2} + \cdots \right), \]
displaying high-energy corrections at order $(H/M)^2$.  The addition of boundary terms such as
\[ S_{\rm boundary} = \int d^3 {\bf x} \frac{\lambda}{M} \partial^i \varphi \partial_i \varphi \]
then modifies the power spectrum at order $H/M$, potentially a more significant observable effect.  Clearly, (\ref{psa}) does not reflect this expectation.  Where did the argument fail?

The answer lies in the assumption of a local operator expansion.  The operators in a UV-complete theory are of course local, and this property is also respected by the low-energy ($p^2/M^2 \ll 1$) expansion of massive Green's functions in static spacetime:
\begin{eqnarray*}
G(x-y) &=& - \int \frac{d^4 p}{(2 \pi)^4} \frac{e^{-ip \cdot (x-y)}}{p^2 - M^2} \\
&\approx&\frac{1}{M^2}  \left[ 1 + \mathcal O \left( \frac{\partial^2}{M^2} \right) \right] \delta^4(x-y) .
\end{eqnarray*}
The Green's (and Wightman's) function used in quasi-de Sitter backgrounds, however, do not yield local effective forms.  Instead, the appropriate expansion parameter is $H/M \ll 1$ which we have seen via the stationary phase approximation is an approximation to energy conservation.  In this expansion, non-local operators are generated which betray the expectation of the small momentum expansion.  This was explicitly shown in \cite{Jackson:2012qp}.

\subsection{Parametrizing Power Spectra}
Using the gauge-conversion in eq. (\ref{gaugeconversion}), the scalar perturbation is then
\begin{equation}
\label{finalps}
P_\zeta(k) = \frac{4 \pi P_\varphi }{M^2_{\rm pl} \epsilon_1}. 
\end{equation}

The advantage of using comoving $k$ to parametrize the power spectrum is that there is a well-defined map to the CMB $l$-modes.  To compare such predictions to observation requires summarizing the power spectrum (\ref{finalps}) by a few parameters which can then be constrained by the data.  The late-time limit of the modefunction is
\[ U_{\bf k} \rightarrow  \frac{H_*}{\sqrt{2 k^3}} \left[ 1 - \epsilon_1 - \left(\epsilon_1 + \frac{\epsilon_2}{2} \right) C - \left( \epsilon_1 + \frac{\epsilon_2}{2} \right) \ln \frac{k}{k_*}  \right] . \]
Here we have used the abbreviation
\begin{eqnarray*}
C &\equiv& - \frac{\Gamma'(3/2)}{ \Gamma(3/2)} - \ln 2 \\
&=&  \gamma + \ln 2 -2
\end{eqnarray*}
where $\gamma$ is the Euler-Mascheroni constant.  Following the example of Martin and Ringeval \cite{Martin:2006rs}, we can then parametrize the scalar power spectrum as:
\begin{equation*}
 \frac{P_{\zeta} (k) }{P_{\zeta 0} (k_*)} = a^{\rm (s)}_0 + a_1^{\rm (s)} \ln \left( \frac{k}{k_*} \right) 
 \end{equation*}
 where
\begin{eqnarray}
\label{psparametrization}
P_{\zeta 0}(k_*) &=& \frac{ H_*^2}{\pi \epsilon_1 M^2_{\rm pl}} , \\
\nonumber
a^{\rm (s)}_0 &=& 1 - 2(C+1) \epsilon_1 - C \epsilon_2 \\
\nonumber
&& \hspace{-0.6in} + \frac{g_1^2 \Lambda^3}{48 \pi M^4 H_*} \left[ 1- \frac{8 \epsilon_1}{3} -  (2  \epsilon_1 + \epsilon_2)  C - 4 \epsilon_1 \ln \left( \frac{4 \Lambda H_*}{M^2} \right) \right], \\
\nonumber
a^{\rm (s)}_1 &=& - 2 \epsilon_1 - \epsilon_2 - \frac{g_1^2 \Lambda^3}{24 \pi M^4 H_*} \left( 6 \epsilon_1 + \epsilon_2 \right).
\end{eqnarray} 
While the search for oscillations in the \emph{WMAP3} data  \cite{Martin:2006rs} failed to show evidence for such high-energy oscillations, the \emph{Planck} satellite data will be of much higher resolution and is worth investigating.  We discuss this matter further in the concluding section.

For the gravity waves, which have no explicit high-energy interactions, the spectrum reduces to the standard slow-roll result.  The late-time limit of the modefunction is
\[ W_{\bf k}  \rightarrow \frac{H_*}{\sqrt{2 k^3}} \left[ 1- \epsilon_1 (C+1) - \epsilon_1 \ln \frac{k}{k_*}  \right]. \]
The parametrization is then
\[ \frac{P_{h} (k) }{P_{h0} (k_*)} = a^{\rm (t)}_0 + a_1^{\rm (t)} \ln \left( \frac{k}{k_*} \right) \]
where
\begin{eqnarray*}
P_{h0}(k_*) &=& \frac{16 H_*^2}{\pi M^2_{\rm pl}} , \\
a^{\rm (t)}_0 &=& 1 - 2(C+1) \epsilon_1, \\
a^{\rm (t)}_1 &=& -2 \epsilon_1.
\end{eqnarray*} 

\section{Observational Possibilities and Conclusion}
In summary, we have developed a technique to explicitly calculate the
generic corrections to the inflaton power spectrum from fundamental
high energy physics.  These can be
connected to microscopic models through a ``generalized'' Boundary
Effective Field Theory as in \cite{Greene:2005aj}. 

Why does the naive effective action of the power spectrum fail? For three reasons:
\begin{enumerate}
\item Correlation functions (such as the power spectrum) must be computed using the in-in formalism, significantly altering the role of classical and quantum fluctuations.
\item The expanding background means that heavy fluctuations can now be excited on-shell.
\item The interactions producing such heavy modes (whether on-shell or virtual) are now localized at specific moments in time, giving rise to effective interaction boundary terms. 
\end{enumerate}

Because there are no distinct features in the power spectrum, any visible new physics will probably need to be found in higher order correlations.  The bispectrum, or 3-point correlation \cite{Komatsu:2009kd}, for this example model will vanish to leading order in slow-roll, but more general models should have the effects of inflaton self-interactions and modified vacua multiplying \cite{Meerburg:2009ys, Meerburg:2009fi}.  Our results justify the conjecture made in \cite{Meerburg:2009fi} that effective actions and modified vacua are expected in tandem.  One distinct hope is that the bispectrum may be oscillatory.

For the trispectrum, or 4-point correlation \cite{Fergusson:2010gn}, little has been done in the way of categorizing such correlations and we cannot give a complete analysis here, but let us summarize a few basic ideas.  Our model has leading-order trispectrum terms of types $\langle {\bar \varphi}^2 \rangle \langle {\bar \varphi}^2 \rangle$ and $ \langle {\bar \varphi}^4 \rangle$.  These include the characteristic oscillations which are now squared, and would display high-energy signatures even if the oscillation frequency $\sim H/M$ were too small to be observed by the power spectrum \footnote{We thank D. Spergel for suggesting the utility of the trispectrum to us.}.

Looking towards complete understanding, further details of the cutoff procedure, renormalization group properties and vacuum selection in an expanding background need to be well-understood.

\section{Acknowledgments}
We would like to thank A.~Ach\'ucarro, R.~Easther, S.~Galli, B.~Greene, D.~Langlois, J.~Martin, P.~D.~Meerburg, S.~Mooij, G.~Palma, C.~Ringeval, D.~Spergel, T.~van der Aalst, B.~Wandelt and especially M. Kleban for helpful discussions and comments on this article.  This research was supported in part by a VIDI and a VICI Innovative Research Incentive Award from the Netherlands Organisation for Scientific Research (NWO), a van Gogh grant  from the NWO, and the Dutch Foundation for Fundamental Research on Matter (FOM).


\begin{thebibliography}{99}

\small
\parskip=0pt plus 2pt

\bibitem{wmap7}
  E.~Komatsu {\it et al.} [WMAP Collaboration],
  arXiv:1001.4538 [astro-ph.CO]; D.~Larson {\it et al.} [WMAP Collaboration],
  arXiv:1001.4635 [astro-ph.CO].

\bibitem{planck}
    [Planck Collaboration],
  ``Planck: The scientific programme,''
  arXiv:astro-ph/0604069.
  

\bibitem{Brandenberger:1999sw}
  R.~H.~Brandenberger,
  arXiv:hep-ph/9910410.
  
 
\bibitem{Niemeyer:2000eh}
  J.~C.~Niemeyer,
  Phys.\ Rev.\  D {\bf 63}, 123502 (2001)
  [arXiv:astro-ph/0005533].


\bibitem{Kempf:2000ac}
  A.~Kempf,
  Phys.\ Rev.\  D {\bf 63}, 083514 (2001)
  [arXiv:astro-ph/0009209].


\bibitem{Niemeyer:2001qe}
  J.~C.~Niemeyer and R.~Parentani,
  Phys.\ Rev.\  D {\bf 64}, 101301 (2001)
  [arXiv:astro-ph/0101451].


\bibitem{Kempf:2001fa}
  A.~Kempf and J.~C.~Niemeyer,
  Phys.\ Rev.\  D {\bf 64}, 103501 (2001)
  [arXiv:astro-ph/0103225].

\bibitem{Martin:2000xs}
  J.~Martin and R.~H.~Brandenberger,
  Phys.\ Rev.\  D {\bf 63}, 123501 (2001)
  [arXiv:hep-th/0005209].


\bibitem{Brandenberger:2000wr}
  R.~H.~Brandenberger and J.~Martin,
  Mod.\ Phys.\ Lett.\  A {\bf 16}, 999 (2001)
  [arXiv:astro-ph/0005432].


\bibitem{Brandenberger:2002hs}
  R.~H.~Brandenberger and J.~Martin,
  Int.\ J.\ Mod.\ Phys.\  A {\bf 17}, 3663 (2002)
 [arXiv:hep-th/0202142].



\bibitem{Martin:2003kp}
  J.~Martin and R.~Brandenberger,
  Phys.\ Rev.\  D {\bf 68}, 063513 (2003)
  [arXiv:hep-th/0305161].


\bibitem{Easther:2001fi}
  R.~Easther, B.~R.~Greene, W.~H.~Kinney and G.~Shiu,
  Phys.\ Rev.\  D {\bf 64}, 103502 (2001)
 [arXiv:hep-th/0104102].


\bibitem{Easther:2001fz}
  R.~Easther, B.~R.~Greene, W.~H.~Kinney and G.~Shiu,
  Phys.\ Rev.\  D {\bf 67}, 063508 (2003)
 [arXiv:hep-th/0110226].

\bibitem{Easther:2002xe}
  R.~Easther, B.~R.~Greene, W.~H.~Kinney and G.~Shiu,
  Phys.\ Rev.\  D {\bf 66}, 023518 (2002)
  [arXiv:hep-th/0204129].


\bibitem{Kaloper:2002uj}
  N.~Kaloper, M.~Kleban, A.~E.~Lawrence and S.~Shenker,
  Phys.\ Rev.\  D {\bf 66}, 123510 (2002)
  [arXiv:hep-th/0201158].


\bibitem{Kaloper:2002cs}
  N.~Kaloper, M.~Kleban, A.~Lawrence, S.~Shenker and L.~Susskind,
  JHEP {\bf 0211}, 037 (2002)
  [arXiv:hep-th/0209231].


\bibitem{Danielsson:2002kx}
  U.~H.~Danielsson,
  Phys.\ Rev.\  D {\bf 66}, 023511 (2002)
 [arXiv:hep-th/0203198].


\bibitem{Danielsson:2002qh}
  U.~H.~Danielsson,
  JHEP {\bf 0207}, 040 (2002)
  [arXiv:hep-th/0205227].


\bibitem{Shankaranarayanan:2002ax}
  S.~Shankaranarayanan,
  Class.\ Quant.\ Grav.\  {\bf 20}, 75 (2003)
  [arXiv:gr-qc/0203060].


\bibitem{Hassan:2002qk}
  S.~F.~Hassan and M.~S.~Sloth,
  Nucl.\ Phys.\  B {\bf 674}, 434 (2003)
  [arXiv:hep-th/0204110].


\bibitem{Goldstein:2002fc}
  K.~Goldstein and D.~A.~Lowe,
  Phys.\ Rev.\  D {\bf 67}, 063502 (2003)
  [arXiv:hep-th/0208167].

\bibitem{Bozza:2003pr}
  V.~Bozza, M.~Giovannini and G.~Veneziano,
  JCAP {\bf 0305}, 001 (2003)
  [arXiv:hep-th/0302184].


\bibitem{Alberghi:2003am}
  G.~L.~Alberghi, R.~Casadio and A.~Tronconi,
  Phys.\ Lett.\  B {\bf 579}, 1 (2004)
  [arXiv:gr-qc/0303035].


\bibitem{Schalm:2004qk}
  K.~Schalm, G.~Shiu and J.~P.~van der Schaar,
  JHEP {\bf 0404}, 076 (2004)
  [arXiv:hep-th/0401164].

\bibitem{porrati}
 M.~Porrati,
  Phys.\ Lett.\  B {\bf 596}, 306 (2004)
  [arXiv:hep-th/0402038].


\bibitem{Porrati:2004dm}
  M.~Porrati,
  arXiv:hep-th/0409210.
  
\bibitem{Hamann:2008yx}
  J.~Hamann, S.~Hannestad, M.~S.~Sloth and Y.~Y.~Y.~Wong,
  JCAP {\bf 0809}, 015 (2008)
  [arXiv:0807.4528 [astro-ph]].
  
\bibitem{Achucarro:2010da}
  A.~Achucarro, J.~O.~Gong, S.~Hardeman, G.~A.~Palma and S.~P.~Patil,
  JCAP {\bf 1101}, 030 (2011)
  [arXiv:1010.3693 [hep-ph]].
  
\bibitem{Starobinsky:2002rp}
  A.~A.~Starobinsky and I.~I.~Tkachev,
  JETP Lett.\  {\bf 76}, 235 (2002)
  [Pisma Zh.\ Eksp.\ Teor.\ Fiz.\  {\bf 76}, 291 (2002)]
  [arXiv:astro-ph/0207572].

\bibitem{Chen:2010ef}
  S.~H.~Chen and J.~B.~Dent,
  arXiv:1012.4811 [astro-ph.CO].

\bibitem{Ashoorioon:2010xg}
 A.~Ashoorioon, G.~Shiu,
 JCAP {\bf 1103}, 025 (2011).
 [arXiv:1012.3392 [astro-ph.CO]].
 
\bibitem{Ashoorioon:2004vm}
 A.~Ashoorioon, A.~Kempf, R.~B.~Mann,
 Phys.\ Rev.\  {\bf D71}, 023503 (2005).
 [astro-ph/0410139].

\bibitem{Bergstrom:2002yd}
  L.~Bergstrom and U.~H.~Danielsson,
  JHEP {\bf 0212}, 038 (2002)
  [arXiv:hep-th/0211006].

\bibitem{Martin:2003sg}
  J.~Martin and C.~Ringeval,
  Phys.\ Rev.\  D {\bf 69}, 083515 (2004)
  [arXiv:astro-ph/0310382].
 
\bibitem{Martin:2004iv}
  J.~Martin and C.~Ringeval,
  Phys.\ Rev.\  D {\bf 69}, 127303 (2004)
  [arXiv:astro-ph/0402609].

\bibitem{Martin:2004yi}
  J.~Martin and C.~Ringeval,
  JCAP {\bf 0501}, 007 (2005)
  [arXiv:hep-ph/0405249].

\bibitem{Easther:2004vq}
  R.~Easther, W.~H.~Kinney and H.~Peiris,
  JCAP {\bf 0505}, 009 (2005)
  [arXiv:astro-ph/0412613].


\bibitem{Greene:2005aj}
  B.~Greene, K.~Schalm, J.~P.~van der Schaar and G.~Shiu,
  [arXiv:astro-ph/0503458].


\bibitem{Easther:2005yr}
  R.~Easther, W.~H.~Kinney and H.~Peiris,
  JCAP {\bf 0508}, 001 (2005)
  [arXiv:astro-ph/0505426].


\bibitem{Spergel:2006hy}
  D.~N.~Spergel {\it et al.}  [WMAP Collaboration],
  Astrophys.\ J.\ Suppl.\  {\bf 170}, 377 (2007)
  [arXiv:astro-ph/0603449].


\bibitem{Cheung:2007st}
  C.~Cheung, P.~Creminelli, A.~L.~Fitzpatrick, J.~Kaplan and L.~Senatore,
  JHEP {\bf 0803}, 014 (2008)
  [arXiv:0709.0293 [hep-th]].

\bibitem{Senatore:2010wk}
  L.~Senatore and M.~Zaldarriaga,
  arXiv:1009.2093 [hep-th].

\bibitem{Weinberg:2005vy}
  S.~Weinberg,
  Phys.\ Rev.\  D {\bf 72}, 043514 (2005)
  [arXiv:hep-th/0506236].


\bibitem{Weinberg:2006ac}
  S.~Weinberg,
  Phys.\ Rev.\  D {\bf 74}, 023508 (2006)
  [arXiv:hep-th/0605244].


\bibitem{Weinberg:2008hq}
  S.~Weinberg,
  Phys.\ Rev.\  D {\bf 77}, 123541 (2008)
  [arXiv:0804.4291 [hep-th]].


\bibitem{Weinberg:2010wq}
  S.~Weinberg,
  Phys.\ Rev.\  D {\bf 83}, 063508 (2011)
  [arXiv:1011.1630 [hep-th]].

\bibitem{Burgess:2002ub}
  C.~P.~Burgess, J.~M.~Cline, F.~Lemieux and R.~Holman,
  JHEP {\bf 0302}, 048 (2003)
  [arXiv:hep-th/0210233].


\bibitem{Burgess:2003zw}
  C.~P.~Burgess, J.~M.~Cline and R.~Holman,
  JCAP {\bf 0310}, 004 (2003)
  [arXiv:hep-th/0306079].


\bibitem{Burgess:2003hw}
  C.~P.~Burgess, J.~M.~Cline, F.~Lemieux and R.~Holman,
  arXiv:astro-ph/0306236.

\bibitem{Jackson:2010cw} 
  M.~G.~Jackson and K.~Schalm,
  Phys.\ Rev.\ Lett.\  {\bf 108}, 111301 (2012)
  [arXiv:1007.0185 [hep-th]].

  
\bibitem{Guth:1980zm}
  A.~H.~Guth,
  Phys.\ Rev.\  D {\bf 23}, 347 (1981).
  
\bibitem{Linde:1981mu}
  A.~D.~Linde,
  Phys.\ Lett.\  B {\bf 108}, 389 (1982).

\bibitem{Albrecht:1982wi}
  A.~Albrecht and P.~J.~Steinhardt,
  Phys.\ Rev.\ Lett.\  {\bf 48}, 1220 (1982).

\bibitem{Linde:1983gd}
  A.~D.~Linde,
  Phys.\ Lett.\  B {\bf 129}, 177 (1983).

\bibitem{Calzetta:1986cq}
  E.~Calzetta and B.~L.~Hu,
  Phys.\ Rev.\  D {\bf 37}, 2878 (1988).
  
\bibitem{Bunch:1978yq}
  T.~S.~Bunch and P.~C.~W.~Davies,
  Proc.\ Roy.\ Soc.\ Lond.\  A {\bf 360}, 117 (1978).

\bibitem{vanderMeulen:2007ah}
  M.~van der Meulen and J.~Smit,
  JCAP {\bf 0711}, 023 (2007)
  [arXiv:0707.0842 [hep-th]].
  
\bibitem{Martin:2006rs}
  J.~Martin and C.~Ringeval,
  JCAP {\bf 0608}, 009 (2006)
  [arXiv:astro-ph/0605367].
  
\bibitem{Einhorn:2003xb}
  M.~B.~Einhorn and F.~Larsen,
  Phys.\ Rev.\  D {\bf 68}, 064002 (2003)
  [arXiv:hep-th/0305056].

\bibitem{Niemack:2010wz}
  M.~D.~Niemack {\it et al.},
  Proc.\ SPIE Int.\ Soc.\ Opt.\ Eng.\  {\bf 7741}, 77411S (2010)
  [arXiv:1006.5049 [astro-ph.IM]].

\bibitem{cmbpol1}
  D.~Baumann {\it et al.}  [CMBPol Study Team Collaboration],
  AIP Conf.\ Proc.\  {\bf 1141}, 3 (2009)
  [arXiv:0811.3911]
  
\bibitem{cmbpol2}
D.~Baumann, M.~G.~Jackson {\it et al.}  [CMBPol Study Team Collaboration],
  AIP Conf.\ Proc.\  {\bf 1141}, 10 (2009)
  [arXiv:0811.3919].
  
\bibitem{Komatsu:2009kd}
  E.~Komatsu {\it et al.},
  arXiv:0902.4759 [astro-ph.CO].
  
\bibitem{Meerburg:2009ys}
  P.~D.~Meerburg, J.~P.~van der Schaar and P.~S.~Corasaniti,
  JCAP {\bf 0905}, 018 (2009)
  [arXiv:0901.4044 [hep-th]].
  
\bibitem{Meerburg:2009fi}
  P.~D.~Meerburg, J.~P.~van der Schaar and M.~G.~Jackson,
  JCAP {\bf 1002}, 001 (2010)
  [arXiv:0910.4986 [hep-th]].
 
\bibitem{Meerburg:2010ks}
  P.~D.~Meerburg,
  arXiv:1006.2771 [astro-ph.CO].

\bibitem{Jackson:2012fu} 
  M.~G.~Jackson and K.~Schalm,
  arXiv:1202.0604 [hep-th].

\bibitem{Fergusson:2010gn}
  J.~R.~Fergusson, D.~M.~Regan and E.~P.~S.~Shellard,
  arXiv:1012.6039 [astro-ph.CO].

\bibitem{Jackson:2012qp} 
  M.~G.~Jackson,
  arXiv:1203.3895 [hep-th].



\end{thebibliography}
\end{document}